\documentclass[12pt,preprint]{aastex}
\usepackage{color}

\newcommand{\cm}{{\rm\,cm}}

\newcommand{\nm}{{\rm\,nm}}
\newcommand{\second}{{\rm\,s}}
\newcommand{\kms}{{\rm\,km\,s^{-1}}}
\newcommand{\cms}{{\rm\,cm\,s^{-1}}}
\newcommand{\au}{{\rm\,AU}}
\newcommand{\AU}{{\au}}
\newcommand{\msun}{{\rm\,M_\odot}}
\newcommand{\Msun}{{\msun}}
\newcommand{\gm}{{\rm\,g}}
\newcommand{\gram}{{\gm}}
\newcommand{\gauss}{{\,\rm G}}
\newcommand{\muG}{{\,\mu\rm G}}
\newcommand{\mum}{{\,\mu\rm m}}

\newcommand{\ct}{\citealt}

\begin{document}
\title{Nonideal MHD Effects and Magnetic Braking Catastrophe in 
Protostellar Disk Formation} 

\author{Zhi-Yun Li\altaffilmark{1,2}, Ruben Krasnopolsky\altaffilmark{2,3}, Hsien Shang\altaffilmark{2,3}}
\altaffiltext{1}{University of Virginia, Astronomy Department, Charlottesville, USA}
\altaffiltext{2}{Academia Sinica, Theoretical Institute for Advanced Research in Astrophysics, Taipei, Taiwan}
\altaffiltext{3}{Academia Sinica, Institute of Astronomy and Astrophysics, Taipei, Taiwan}

\shortauthors{{\sc Li, Krasnopolsky, and Shang}}
\shorttitle{{\sc Non-ideal MHD Effects in Disk Formation}}

\begin{abstract}
Dense, star-forming, cores of molecular clouds are observed 
to be significantly magnetized. A realistic magnetic field 
of moderate strength has been shown to suppress, through 
catastrophic magnetic 
braking, the formation of a rotationally supported disk 
during the protostellar accretion 
phase of low-mass star formation in the ideal MHD limit. We 
address, through 2D (axisymmetric) simulations, the 
question of whether realistic levels of nonideal effects, computed with a 
simplified chemical network including dust grains, can weaken the 
magnetic braking enough to enable a rotationally supported disk 
to form. We find that ambipolar diffusion, the dominant nonideal 
MHD effect over most of the density range relevant to disk 
formation, does not enable disk formation, at least in 2D. The 
reason is that ambipolar diffusion allows the magnetic flux that
would be dragged into the central stellar object in the ideal 
MHD limit to pile up instead in a small circumstellar region, 
where the magnetic field strength (and thus the braking efficiency) 
is greatly enhanced. We also find that, on the scale of tens of 
AU or more, a realistic level of Ohmic dissipation does not weaken 
the magnetic braking enough for a rotationally supported disk 
to form, either by itself or in combination with ambipolar 
diffusion. The Hall effect, the least explored of these three 
nonideal MHD effects, can spin up the material close to the 
central object to a significant, supersonic rotation speed, 
even when the core is initially non-rotating, although the 
spun-up material remains too sub-Keplerian to form a rotationally 
supported disk. The problem of catastrophic magnetic braking 
that prevents disk formation in dense cores magnetized to 
realistic levels remains unresolved. Possible resolutions 
of this problem are discussed. 

\end{abstract}

\section{Introduction}  
\label{intro}

The formation and early evolution
of disks is a long-standing fundamental problem in star formation. 
Early work in the field had concentrated on the simpler problem of 
disk formation from the collapse of a rotating dense core in the 
absence of a magnetic field, as reviewed in \citet{Bodenheimer1995} and 
\citet{Boss1998}. Dense star-forming cores are observed to be 
significantly magnetized, however. There is increasing theoretical 
evidence that disk formation is greatly modified, perhaps even 
suppressed, by a dynamically important magnetic field.  

The most comprehensive measurements of the magnetic field strength 
in dense cores of low-mass star formation come from \citet{TrolandCrutcher2008}.
They carried out an OH Zeeman survey of 
a sample of nearby dark cloud cores, probing densities of order 
$10^3$--$10^4\cm^{-3}$. The inferred mean value for the dimensionless 
mass-to-flux ratio, relative to the critical value $(2\pi
G^{1/2})^{-1}$ (\ct{NakanoNakamura1978}, \ct{ShuLi1997}), is   
$\lambda_{los}\approx 4.8\pm 0.4$. It was obtained from the 
measured line-of-sight component of the magnetic field $B_{los}$
without any geometric correction. Correcting for geometric effects
statistically would lower the mass-to-flux ratio by a factor of 2--3
\citep{Shu_1999}, bringing the mean value of the intrinsic
mass-to-flux ratio to a few, i.e., $\lambda \sim$ 2--3. Such values 
of $\lambda$ are naturally produced in the scenario of ambipolar
diffusion-regulated dense core formation in strongly magnetized 
(magnetically subcritical) clouds, with or without the assistance 
of supersonic turbulence 
(e.g., \ct{LizanoShu1989}; \ct{BasuMouschovias1994};
\ct{NakamuraLi2005}; \ct{KudohBasu2011}). 
It may also be consistent with the scenario of dense core 
formation from turbulence compression in more weakly magnetized 
background clouds, since the core is expected to be more strongly 
magnetized relative to its mass (i.e., lower $\lambda$) than the 
cloud as a whole (\ct{TilleyPudritz2005}; \ct{Dib_2007}). Well 
ordered magnetic fields are also inferred from polarization maps 
of dust continuum emission, both on the core scale \citep{Ward-Thompson_2000}
and smaller (e.g., \ct{Girart_2006}). We therefore 
expect the dense cores to be rather strongly magnetized based on 
both observational data and core formation theories.      

A moderately strong magnetic field can suppress disk formation in 
the ideal MHD limit. This was first demonstrated in \citet{Allen_2003},
who carried out 2D (axisymmetric) simulations of the 
collapse of rotating cores magnetized to a level of $\lambda \leq 
10$. The basic reason for disk suppression is that, in the ideal 
MHD limit, flux freezing allows the infalling material to drag a 
finite amount of magnetic flux into the central object, creating
a split magnetic monopole whose (poloidal) field strength 
increases rapidly with decreasing radius (as $r^{-2}$, \ct{Galli_2006};
see their Fig.\ 1). The increased field strength close 
to the central object, coupled with a long magnetic lever arm from 
severe equatorial pinching of (poloidal) field lines, is responsible for 
the catastrophic magnetic braking that suppresses the formation 
of a rotationally supported
disk. This magnetic braking catastrophe was confirmed 
numerically by \citet{MellonLi2008} and \citet{HennebelleFromang2008} 
using, respectively, 2D and 3D simulations (see also \ct{PriceBate2007}),
at least when the magnetic and rotation axes 
are aligned (see, however, \ct{Machida_2010} and discussion 
in \S{}\ref{discussion}). 

When the magnetic and rotation axes are misaligned, \citet{HennebelleCiardi2009}
found in their AMR MHD simulations that magnetic 
braking efficiency is reduced relative to the aligned case. A
potential concern is that the misalignment increases the flow 
complexity, which may enhance the numerical magnetic diffusion 
that is considerable on small scales (see a nice discussion in 
\S{}5.3 of \ct{Hennebelle_2011}). A more obvious possibility for 
avoiding the magnetic braking catastrophe is to relax the 
ideal MHD approximation. Since dense cores are known to be 
lightly ionized \citep{BerginTafalla2007}, non-ideal MHD effects 
(including ambipolar diffusion, Hall effect and Ohmic dissipation, 
e.g., \ct{Nakano_2002}) are to be expected. The first non-ideal 
MHD effect considered in this context was Ohmic dissipation. \citet{Shu_2006}
suggested that in order for Ohmic dissipation to 
weaken the field strength over a large enough region so that a 
large-scale rotationally supported disk of tens of AUs or more can potentially 
form, the resistivity must be at least one order of magnitude 
above the classical (microscopic) value. This suggestion was
confirmed by \citet{Krasnopolsky_2010}, who found that large, 
$100\AU$ scale disks can indeed form, as long as the resistivity 
is enhanced by a large enough factor, to a value of order 
$10^{19}\cm^2\second^{-1}$ or more. \citet{Machida_2010} 
carried out core collapse calculations including a distribution 
of resistivity with density and temperature (from a fit to the 
resistivities computed in \ct{Nakano_2002}) and found that, 
even with just the 
classical resistivity, a small rotationally supported disk can form 
at the beginning of the protostellar accretion phase (see also 
\ct{DappBasu2010}) and grow to larger, 100-AU scales at later 
times. Part of the apparent discrepancy between \citet{Machida_2010} 
and \citet{Krasnopolsky_2010} may be due to different simulation 
setup. Another difference may be in the level of numerical 
magnetic diffusivity (see discussion in \S{}\ref{discussion}).
Further investigation is needed to clarify the situation in the 
limiting case that includes only Ohmic dissipation. 

Ohmic dissipation is important at high densities (above $\sim 
10^{11}\cm^{-3}$, \ct{Nakano_2002}; see also \ct{KunzMouschovias2010}).
Before reaching 
such densities, dense core material must evolve through lower 
densities, where the Hall effect and especially ambipolar 
diffusion dominate over Ohmic dissipation. The effect of ambipolar 
diffusion on disk formation was investigated semi-analytically 
by \citet{KrasnopolskyKonigl2002},
and numerically by \citeauthor{MellonLi2009}(\citeyear{MellonLi2009};
see also \ct{DuffinPudritz2009}
and \ct{HoskingWhitworth2004}). Compared to 
the ideal MHD case, a new ingredient is the ambipolar 
diffusion-induced accretion shock, driven by the magnetic flux 
left behind by the material that has gone into the central 
object (\ct{LiMcKee1996}; \ct{CiolekKonigl1998}; \ct{Contopoulos_1998};
\ct{TassisMouschovias2007}). \citet{KrasnopolskyKonigl2002}
demonstrated using a 1D semi-analytic model that the 
strong magnetic field piled up inside the ambipolar diffusion 
(AD) shock can in principle brake the post-shock material efficiently.
\citet{MellonLi2009} showed through 2D (axisymmetric) simulations 
that this is indeed the case. They started their calculations 
from a self-similar rotating, magnetized isothermal toroid 
\citep{Allen_2003}. A power-law dependence on the neutral 
density is assumed for the ion density, so that 
the subsequent collapse remains self-similar, 
even in the presence of ambipolar diffusion. The self-similarity 
provides a powerful check on the validity of the numerically 
obtained collapse solutions. It imposes, however, strong 
restrictions on both the initial core properties and 
charge densities. 

One objective of the present paper is to investigate the role of 
ambipolar diffusion in disk formation without the restrictive 
simplifications made in \citet{MellonLi2009}. 
We do this by following both the pre-stellar evolution of the
rotating, magnetized dense core and the protostellar mass 
accretion phase after a central stellar object has formed and 
by computing the charge densities self-consistently 
using a simplified chemical network that includes dust grains 
(\ct{Nishi_1991}; \ct{Nakano_2002}); they can affect the 
magnetic diffusivities greatly (e.g., \ct{WardleNg1999}).  
Another objective is to extend
\citeauthor{MellonLi2009}'s (\citeyear{MellonLi2009}) calculations 
by including both 
Ohmic dissipation and Hall effect in addition to ambipolar 
diffusion. The Hall effect was explored previously in the 
context of disk-driven outflows (\ct{WardleKonigl1993}; see also 
\ct{Konigl_2010}) and 
accretion disk dynamics \citep{SanoStone2002};
it is only starting to be explored in the context of core 
collapse and disk formation (\ct{Krasnopolsky_2011};
\ct{Braiding2011}).

We find that, on scales greater 
than $10^{14}\cm$ (or $6.7\AU$) that we are able to resolve in 
our non-ideal MHD calculations, no rotationally supported disks 
form in dense cores with a moderately strong magnetic field  
(with $\lambda \sim$ several), largely because of the excessive 
braking due to the magnetic field trapped interior to the 
AD shock. On these scales, Ohmic dissipation affects the flow 
dynamics relatively little. The Hall effect, on the other 
hand, can torque up the spun-down post AD-shock material 
to significant, supersonic, rotation speeds. The rotation 
speeds remain well below Keplerian and rotationally supported 
disks are not formed. Our non-ideal MHD calculations re-enforce 
the idea that disk formation is difficult in the presence of a 
moderately strong magnetic field, at least in 2D (assuming 
axisymmetry). We discuss possible ways to get around this 
difficulty in the discussion section (\S{}\ref{discussion}). 

The rest of the paper is organized as follows. In \S{}\ref{setup}, 
we describe the problem setup, including the initial and boundary 
conditions, the nonideal MHD code used in this work, and
the computation of charge densities including dust grains. Disk 
suppression by a moderately strong magnetic field under a range 
of realistic conditions is demonstrated in \S{}\ref{suppression}. 
We discuss potential disk formation in the case of weaker 
magnetic fields in \S{}\ref{weakfield} and the spin up of an  
initially non-rotating collapsing envelope by the Hall effect 
in \S{}\ref{PureHall}. Our main results are summarized \S{}\ref{summary}.

\section{Problem Setup}
\label{setup}

\subsection{Initial and Boundary Conditions}
\label{initialBC}

Low-mass pre-stellar cores in nearby star-forming regions are 
observed to have relatively simple dynamical structures 
\citep{BerginTafalla2007}. We idealize such cores as initially
uniform spheres of $R=10^{17}\cm$ in mass and $1\msun$ 
in mass (see \ct{HennebelleFromang2008} for a similar setup). 
The initial core mass density is thus $\rho_0=4.77\times 
10^{-19}\gram\cm^{-3}$, corresponding to a volume density for 
molecular hydrogen $n_{H_2} = 10^5\cm^{-3}$ (assuming 10 
hydrogen nuclei for each He), and a free fall time $t_{ff}
=3\times 10^{12}\second$. A simple isothermal equation of state 
is adopted below a critical density $\rho=10^{-13}\gram\cm^{-3}$ 
(with a sound speed $a=0.2\kms$) and  $P\propto \rho^{7/5}$ 
at densities above (e.g., \ct{MasunagaInutsuka2000}). 
The ratio of thermal 
to gravitational binding energy is $\alpha = 2.5 R a^2/(GM)=0.75$. On 
this uniform sphere, we impose a uniform magnetic field $B_0$ at the 
beginning of the simulation. We choose a fiducial value for $B_0$ of 
$10^{-5}$ in the Lorentz-Heaviside units that are convenient for the 
Zeus family of codes. It corresponds to $B_0=\sqrt{4\pi}\times 
10^{-5}\gauss = 35.4\muG$ in Gaussian CGS units. The ratio of magnetic 
to gravitational binding energy is $\gamma = 0.13$. Another way 
to characterize the strength of the magnetic field is through the 
dimensionless mass-to-flux ratio $\lambda$, in units of the critical 
value $(2\pi G^{1/2})^{-1}$. For the core as a whole, the mass-to-flux
ratio is $\lambda = 2.92$. On the central flux tube that 
passes through the origin, the mass-to-flux ratio is $\lambda_
c = 4.38$, higher than the global value by $50\%$. These 
fiducial values of mass-to-flux ratios are somewhat larger than the
values of $\lambda \sim$ 2--3 obtained in models of ambipolar-diffusion 
driven core formation out of magnetically subcritical background 
clouds, with (\ct{NakamuraLi2005}; \ct{KudohBasu2011}) or without 
(\ct{LizanoShu1989}; \ct{BasuMouschovias1994}) turbulent compression. 
They are consistent with the mean value of $\lambda$ inferred by 
\citet{TrolandCrutcher2008} after correcting for geometric effects 
(see \S{}1).

One may argue that the adopted fiducial field strength of $B_0 = 
35.4\muG$ at a number density $n_{H_2}=10^5\cm^{-3}$ is too 
high, because such a field
strength is rarely measured in nearby regions of low-mass star 
formation (such as the Taurus clouds) directly using OH. However,  
the OH emission is dominated by relatively low density envelopes 
of dense cores \citep{TrolandCrutcher2008}.
If the core material was condensed out of a more diffuse 
gas, its field strength would be lower 
before condensation. For example, at densities of order $n_{H_2}
\sim 10^{3.5}\cm^{-3}$ (probed by typical OH observations; 
\ct{Crutcher_2010}), the pre-condensation field strength of 
our model core would be only $\sim 3.5\muG$ (if the condensation 
occurs more or less isotropically under flux-freezing condition, as 
it should be since the core is substantially magnetically 
supercritical). It is lower than the medium value of $\sim 6\muG$ 
inferred for the {\it atomic} cold neutral medium
\citep{HeilesTroland2005}.
If anything, our adopted fiducial field strength may be on the low
side. 
Nevertheless, we will consider some smaller values for $B_0$ 
as well, in view of the possibility that there may be a broad 
distribution of field strengths at a given density \citep{Crutcher_2010}
and the fact that the majority of stars are formed 
in clusters where the magnetic field is less well observed. 

\begin{figure}
\epsscale{1.2}
\plottwo{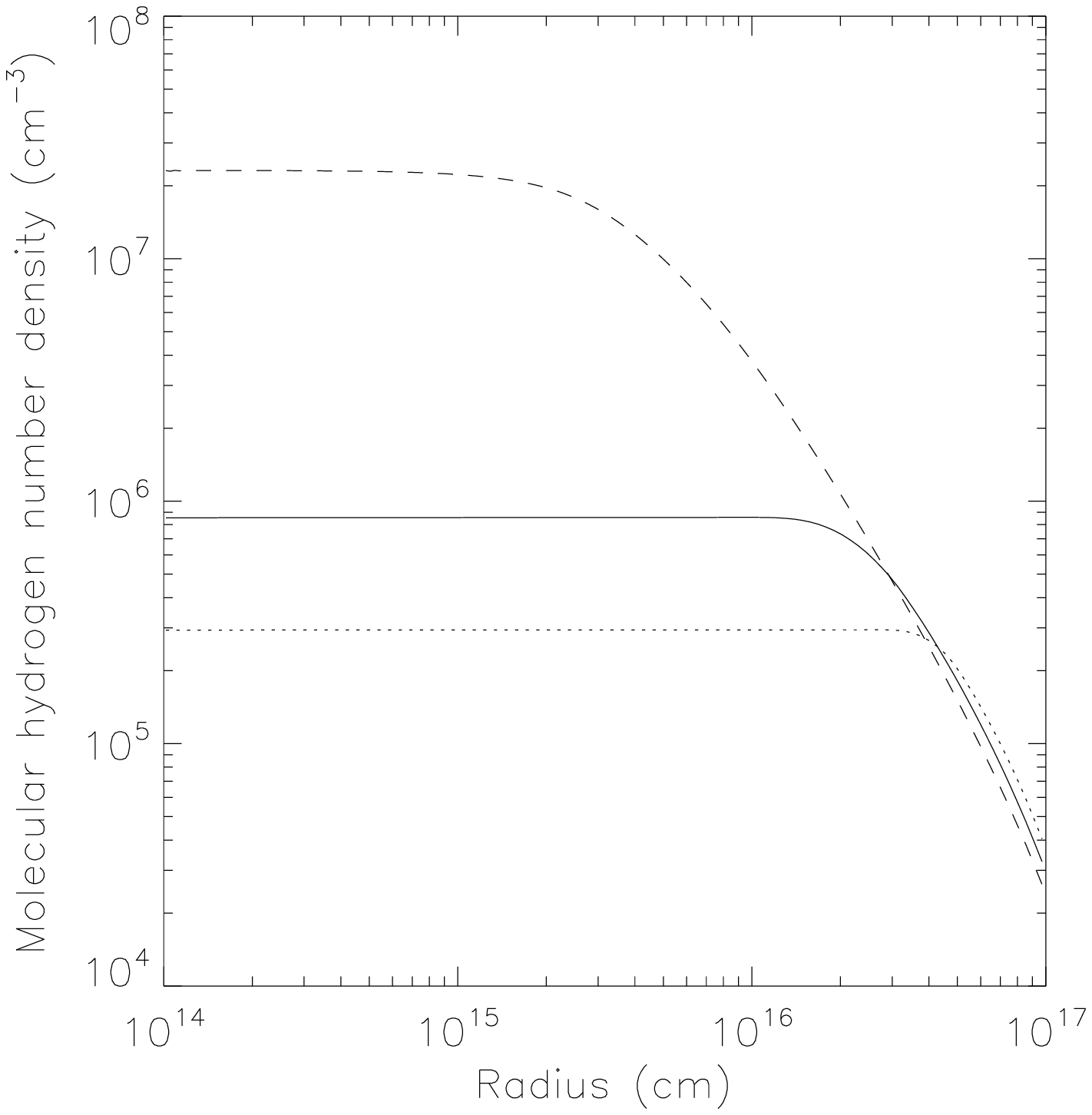}{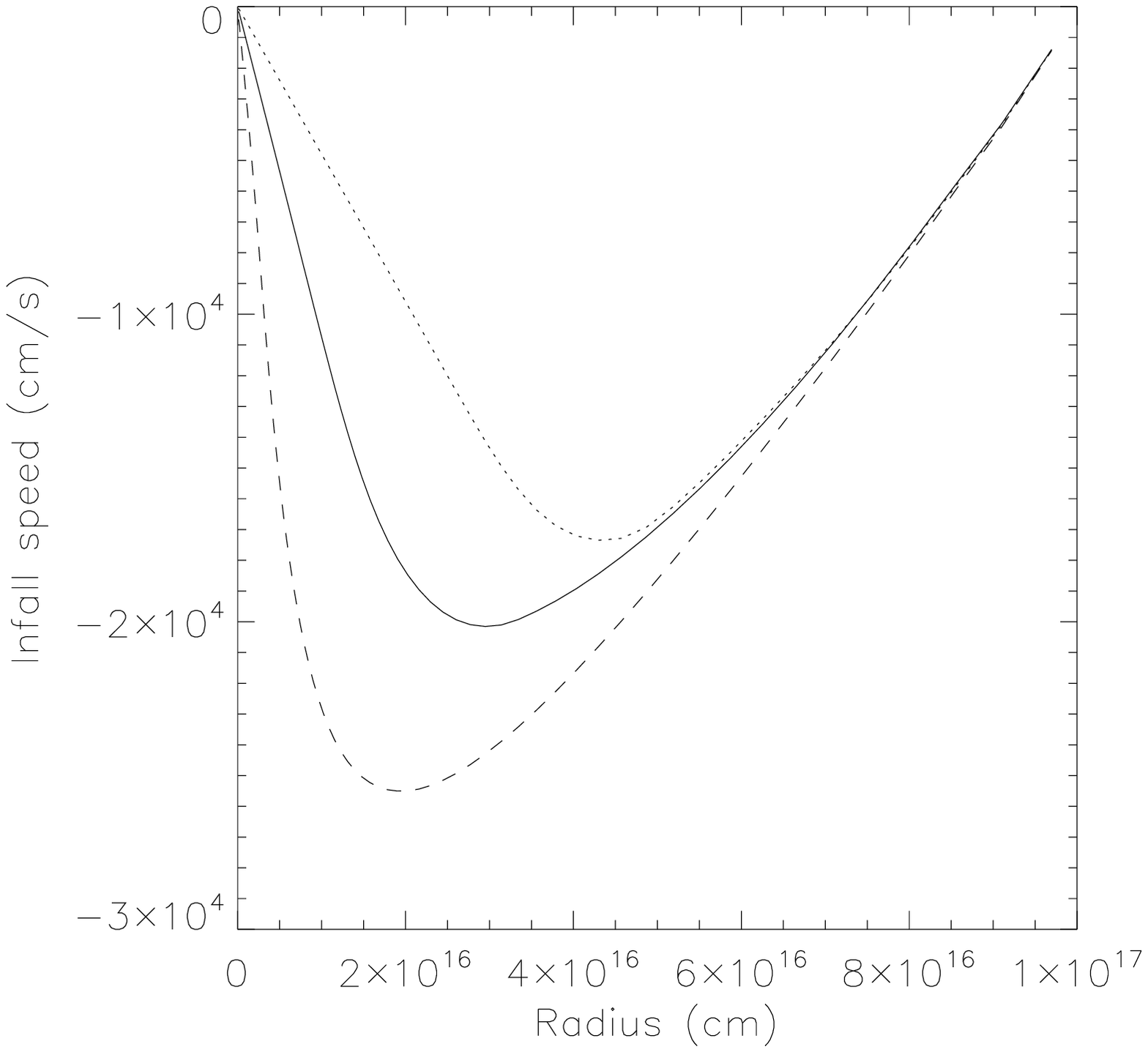}
\caption{Distributions of the density (left panel) and infall 
 velocity (right) at three times $2\times 10^{12}$ (dotted line), 
$2.5\times 10^{12}$ (solid) and $3\times 10^{12}\second$ (dashed) 
for the collapse of a non-rotating, non-magnetize dense core.} 
\label{condensation}
\end{figure}

One may also be concerned that our adopted initial density
distribution is uniform, whereas the observed pre-stellar 
cores are centrally condensed \citep{BerginTafalla2007}. 
However, the density distribution changes with time, evolving 
through a series of centrally condensed configurations with 
different degrees of central-to-edge contrast, as seen in 
Fig.\ \ref{condensation} in the absence of magnetic field and 
rotation.  
For example, the central density $n_{H_2}\sim 10^6\cm^{-3}$ at 
time $2.5\times 10^{12}\second$ is close to that inferred for the well 
studied pre-stellar core L1544 \citep{Ward-Thompson_1999}, 
although the maximum infall speed at this time is somewhat higher 
than the observationally inferred value, which is roughly 
$10^{4}\cms$ \citep{Tafalla_1998}. The infall speed is 
reduced, however, by both rotation and especially magnetic fields 
that are included in the majority of our calculations. Indeed, 
the infall is approximately half-sonic for our reference model 
(see Table 1) when the central density $n_{H_2}\sim 10^6\cm^{-3}$, in agreement 
with observations.

An alternative approach would be to start with a centrally condensed 
static configuration that mimics the observed pre-stellar 
density profile, with a uniform magnetic field imposed at the
beginning (e.g., \ct{Machida_2007}). Implicit in this approach is
the assumption that there is strong evolution in the density 
profile of the core but no corresponding evolution in its magnetic
field distribution. For our purpose of 
studying the efficiency of magnetic braking, it has the drawback 
that the central flux tubes are loaded with much more mass than 
those in the outer part. 
It would lead to a weaker magnetic braking in the early phase 
of protostellar collapse that, in our view, is hard to justify 
physically. 

The rotating profile of dense cores is not well constrained 
by observations. For simplicity, we adopt for our initially 
uniform core a solid body rotation, with a fiducial angular 
speed $\Omega = 10^{-13}\second^{-1}$. It corresponds 
to a ratio of rotational to gravitational binding energy
$\beta = 0.025$, which is typical of the values inferred for 
NH$_3$ cores based on the velocity gradient observed across 
the cores \citep{Goodman_1993}; the inferred $\beta$  
has a considerable range, which motivates us to consider 
other values of $\beta$ as well.

As in \citet{MellonLi2008,MellonLi2009}, we adopt a spherical polar 
coordinate system $(r, \theta, \phi)$ for our axisymmetric 
simulation. The inner boundary is 
set at $10^{14}\cm$ (or $6.7\AU$) and the outer boundary at
$10^{17}\cm$. Even though our inner radius is relatively 
large compared with other collapse studies, such as Machida 
et al. (2010) who set the sink particle size to $1\AU$, our
smallest cell size is only $0.2\AU$, which ensures that the 
flow dynamics near the inner boundary is well resolved. The
high resolution is particularly important for minimizing the
numerical magnetic diffusion that may affect the trapping 
of magnetic flux at small radii, which lies at the heart 
of efficient magnetic braking and disk suppression. The 
standard outflow boundary conditions are enforced at both 
the inner and outer boundaries. Matter that crosses the inner 
boundary is collected at the origin. The central point mass 
interacts with the matter in the computation domain through 
gravity. We use a computational grid 
of $120\times 90$ that is non-uniform in the $r$-direction, with 
a spacing $\Delta r=0.2\AU$ next to the inner boundary. The  
spacing increases outward by a constant factor $\sim 1.0647$ from one 
cell to the next. The grid is uniform in the $\theta$-direction. 
 
\subsection{Induction Equation and Non-Ideal MHD Code}

At the heart of our non-ideal MHD core collapse problem lies the
induction equation
\begin{equation}
{\partial {\bf B}\over \partial t }= \nabla \times ({\bf v}\times {\bf B}) 
  - \nabla \times [\eta_O (\nabla \times {\bf B})] 
  - \nabla \times \left\{ \eta_H [(\nabla\times {\bf B})\times {{\bf
        B}\over B}]\right\} 
  - \nabla \times \left\{ \eta_A {{\bf B}\over B}  \times [(\nabla \times
        {\bf B})\times {{\bf B} \over B}] \right\}
\end{equation}
where the Ohmic, Hall and ambipolar diffusivities are related to the
electric conductivity parallel to the field line $\sigma_\parallel$,
Pedersen and Hall conductivities $\sigma_P$ and $\sigma_H$ through 
\citep{Nakano_2002}
\begin{equation}
\eta_O = {c^2\over 4\pi \sigma_\parallel}; \ \ \ \ \eta_H= {c^2 \over
  4\pi} {\sigma_H \over \sigma_P^2+\sigma_H^2 }; 
  \ \ \ \ \eta_A= {c^2\over 4\pi} 
  \left({\sigma_P\over \sigma_P^2 +\sigma_H^2} -{1\over
  \sigma_\parallel}\right).  
\end{equation}
These diffusivities will be discussed further in the next subsection. 
Here, we describe briefly our numerical treatment of the three 
non-ideal MHD terms. 

Our code, dubbed ``ZeusTW,'' was derived from the ideal MHD code 
Zeus3D \citep{Clarke_1994}. 
It treats the Ohmic term in the 
induction equation (1) using an algorithm based on \citeauthor{Fleming_2000}
(\citeyear{Fleming_2000}; see also \ct{Krasnopolsky_2010}). Ambipolar diffusion 
was treated using the explicit method described in \citet{MacLow_1995},
as in \citet{MellonLi2009}. The magnetic field is evolved 
using a velocity that is the sum of the bulk neutral velocity ${\bf v}$ 
and a ``drift'' velocity defined as 
\begin{equation}
\Delta {\bf v}_{AD} =  {\eta_A \over B^2} (\nabla \times {\bf B}) 
\times {\bf B}.
\end{equation}
In the widely discussed limit where ambipolar diffusion is the 
dominant non-ideal effect and ions are well tied to the magnetic 
field, the velocity $\Delta {\bf v}_{AD}$ can be interpreted as the 
ion-neutral drift velocity. In the more general 
case that we are studying, this is not necessarily the case. 
Nevertheless, we will define an ``effective ion velocity'' 
\begin{equation}
{\bf v}_{i,{\rm{eff}}} \equiv \Delta {\bf v}_{AD} + {\bf v},
\end{equation}
to make contact with previous work. It provides a useful measure 
of the effect of ambipolar diffusion even in the presence of 
other non-ideal MHD effects. Lastly, our treatment of the 
Hall term was based on \citet{SanoStone2002} and
\citeauthor{Huba2003} (\citeyear{Huba2003}, see 
\ct{Krasnopolsky_2011} for detail). 
The time steps for evolving the ambipolar diffusion and Hall terms in 
the induction equation are particularly stringent near the 
polar axes, where the magnetic field is strong but the density 
is often low, due to either gravitational collapse along the 
field lines or outflow. In some cases, a floor is imposed on 
the time step, by decreasing the ambipolar or Hall diffusivity 
in a small volume or increasing the density in an evacuated 
region to limit the Alfv{\'e}n speed. We have verified that the floor has 
little effect on the flow dynamics.     

\subsection{Charge Densities and Magnetic Diffusivities}
\label{s:charge} 

The magnetic diffusivities depend on the densities of charged
particles, including molecular and atomic ions, electrons,
and charged dust grains. We will follow Nakano and collaborators 
in computing the charge densities (e.g., \ct{Nakano_2002}; 
\ct{Nishi_1991}), 
using a simplified chemical network and simple prescriptions 
for grain size distribution. The network includes neutral species
H$_2$, He, CO, O, O$_2$ and other heavy metals (denoted collectively 
by ``M'') and charged species $e^-$, H$^+$, He$^+$, C$^+$, M$^+$, 
H$_3^+$, and other molecular ions (denoted collectively by 
``$m^+$''), as well as neutral and positively and negatively 
charged dust grains (see \ct{Nishi_1991} for detail). The grain  
size distribution 
in dense cores of molecular clouds is relatively unconstrained 
observationally. For illustration, we will consider the 
standard MRN distribution $(dn/da)\propto a^{-3.5}$ with the grain 
size $a$ between $5\nm  < a < 250\nm$ \citep{Mathis_1977} 
and a grain mass that is $1\%$ of the total. It is likely, however, 
for the grains to grow substantially in dense cores of molecular 
clouds; the MRN distribution, which is more appropriate for
diffuse clouds, may be regarded as the starting point for grain 
growth in dense cores. Direct evidence for grain growth comes 
from the Spitzer detection of the so-called ``coreshine'' 
\citep{Pagani_2010}, which indicates that at least some grains 
have grown to micron-size. To illustrate the effects of grain 
growth, we will also consider an opposite limit where the grains 
have a single, large, size $a=1\mum$ (denoted LG distribution). 
The MRN and LG distributions should bracket the real situation 
where some grain growth is expected.   

\begin{figure}
\epsscale{0.45}
\plotone{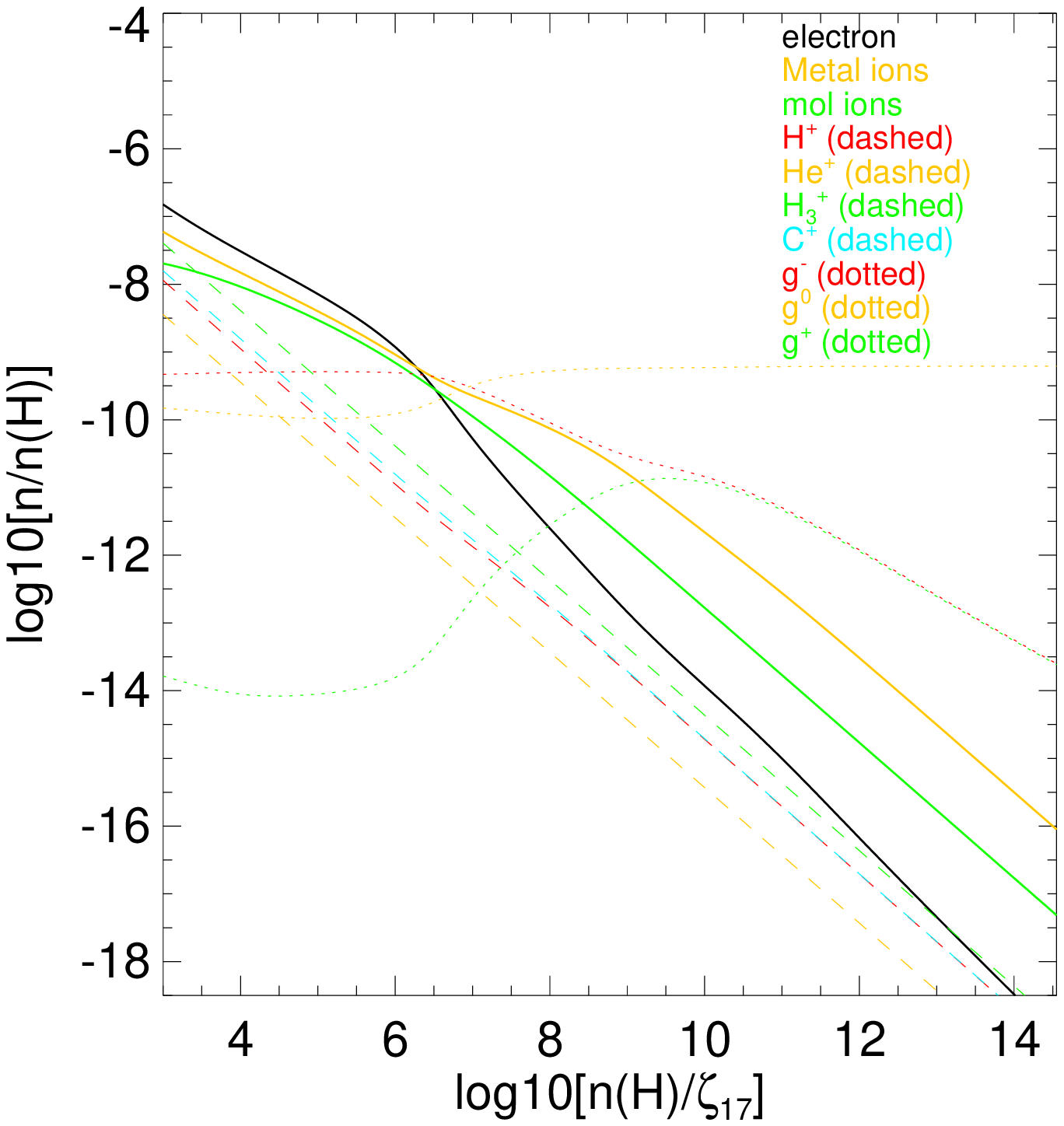}
\plotone{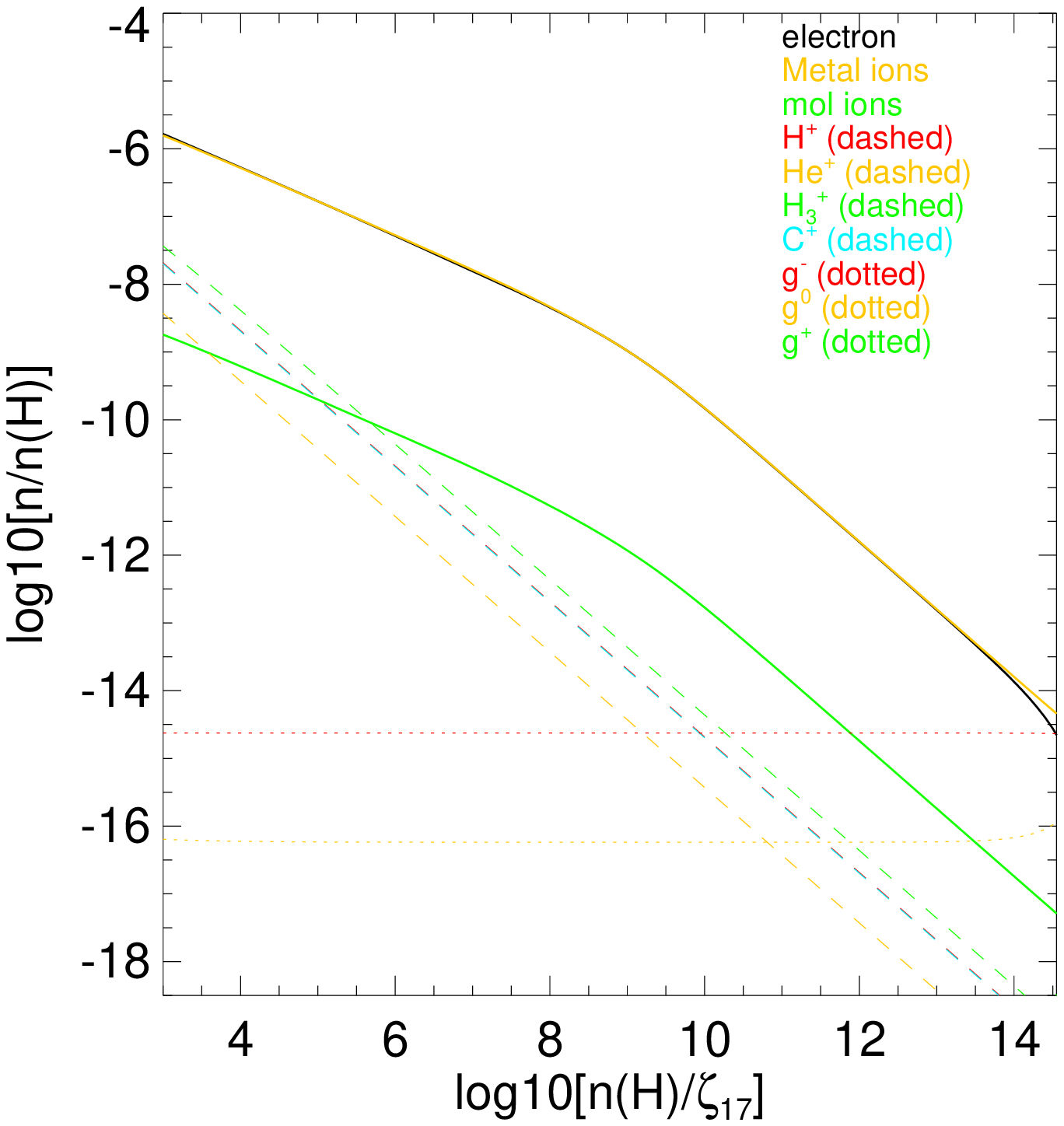}
\caption{The number densities of charges for the standard MRN
  grain size distribution (left panel) and large, $1\mum$ 
grains (right). } 
\label{f:charge}
\end{figure}

Fig.\ \ref{f:charge} plots the fractional abundances (relative to 
the number density of hydrogen nuclei $n_H$) of the charged atomic 
and molecular species and dust grains as a function of 
$n_H$ (related to mass density through $\rho = 2.34\times 10^{-24} 
n_H$), with the cosmic rate ionization rate $\zeta$ normalized to 
the standard value $10^{-17}\second^{-1}$; the normalized rate is
denoted by $\zeta_{17} = \zeta/(10^{-17}\second^{-1})$. In the 
(simpler) large grain (LG) case, the dominant charges are metal 
ions ($M^+$) and electrons ($e^-$) over the whole density range 
of interest to us. 
In the MRN case, the charge densities are lower compared to the 
LG case, because of a large amount of small grains, which provide
a large total surface area for ions and electrons to recombine on. 
The metal ions and electrons remain the most abundant charges at 
low densities (below $\sim 10^6\cm^{-3}$). At higher densities, 
negatively and positively charged (small) grains become more 
dominant. The MRN case was computed with 10 size bins equally spaced 
logarithmically, following \citet{Nishi_1991}. When 20 bins are used,
the charge densities change by less than $3\%$ over the density range
that spans more than 10 orders of magnitude.    
 
\begin{figure}
\epsscale{0.45}
\plotone{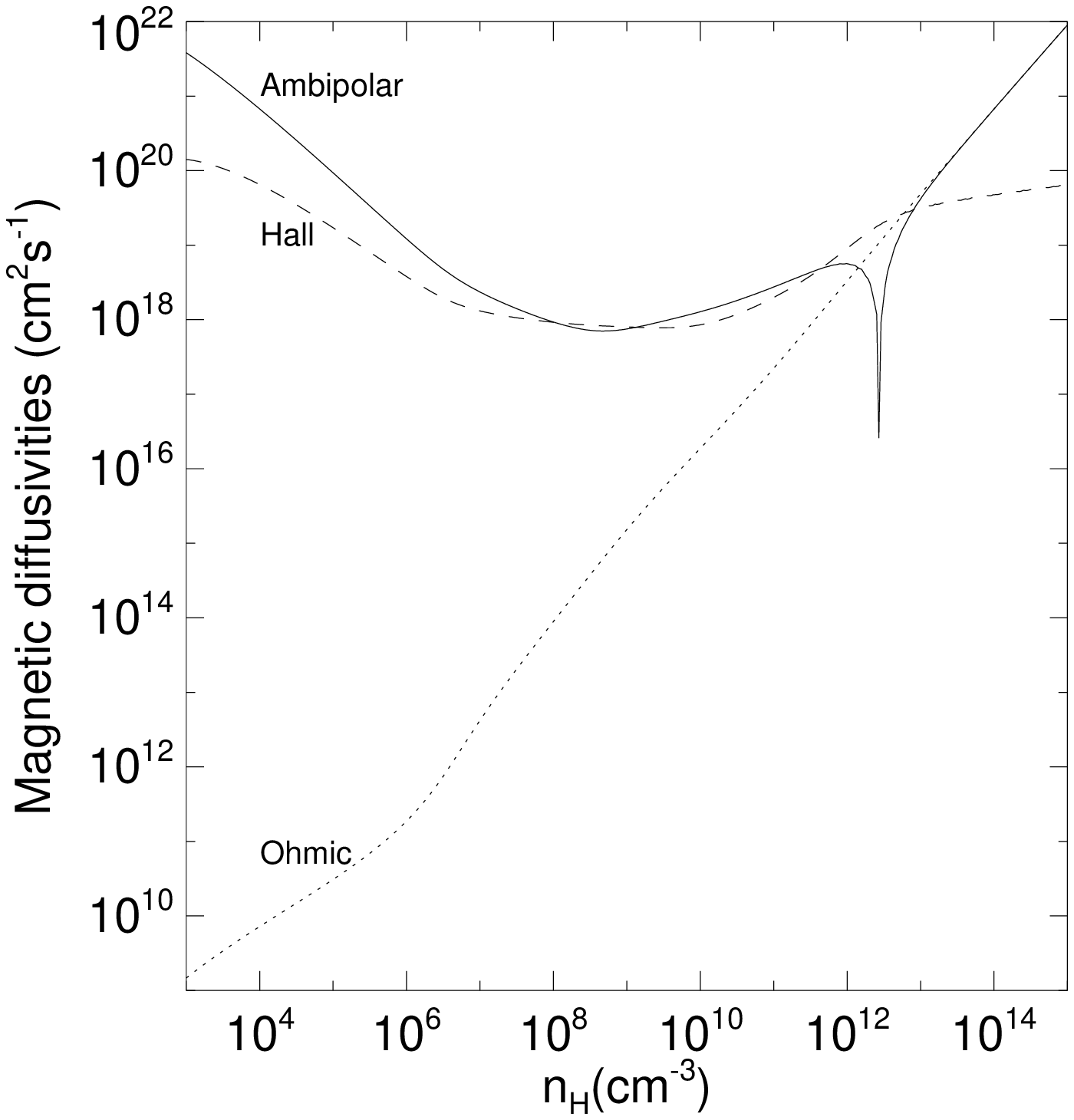}
\plotone{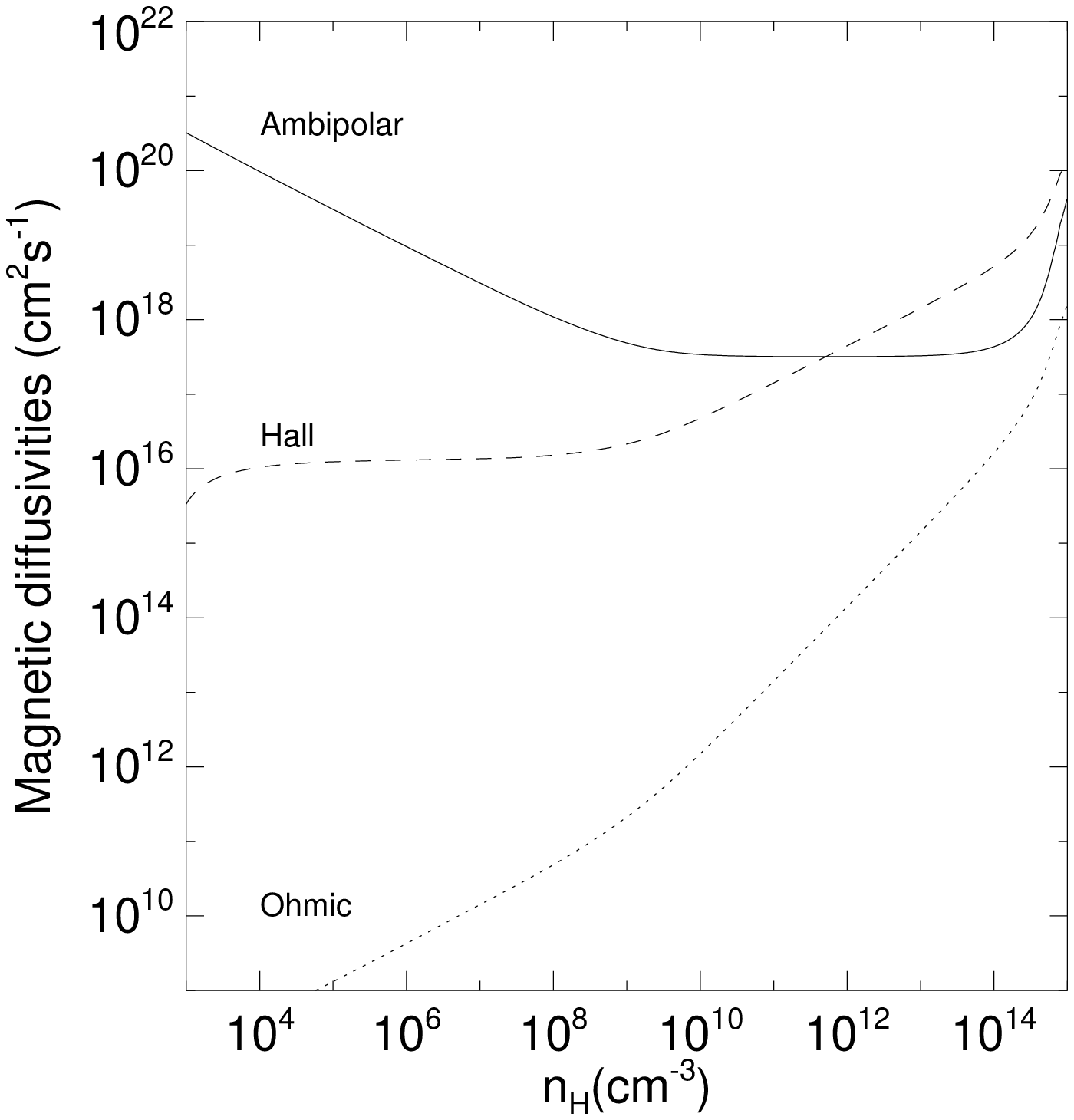}
\caption{The absolute values of the computed magnetic diffusivities
  for the MRN (left panel) and large grain (right panel) cases, for 
the illustrative magnetic field-density relation given in eq.\ (5). For the MRN
  case, the Hall diffusivity $\eta_H$ is negative everywhere and
the ambipolar diffusivity $\eta_A$ changes from positive to negative 
as the density $n_H$ increases above $\sim 10^{12}\cm^{-3}$. } 
\label{Diffusivities}
\end{figure}

To obtain the magnetic diffusivities in the induction equation, we 
need not only charge densities, but also magnetic field strength. 
The field strength will be computed self-consistently in our MHD
simulations. The rough magnitude of the diffusivities can be 
illustrated using e.g. the field strength-density relation
\begin{equation}
B= 1.43 \times 10^{-7} n_H^{1/2}
\end{equation}
assumed in \citet{Nakano_2002}, relevant for a magnetically supported 
sheet in hydrostatic equilibrium along the field lines. The computed 
diffusivities are shown in Fig.\ \ref{Diffusivities}. As expected, 
the ambipolar diffusivity dominates at relatively low densities 
for both MRN and LG cases. For the MRN case, which has a large amount 
of small grains, the Hall diffusivity becomes comparable to the 
ambipolar diffusivity over a wide range of density ($\sim 10^8$ -- 
$10^{12}\cm^{-3}$). It has a negative value, dominated by the
contribution from negatively charged grains, although the contribution
from positively charged grains approaches (from below) that from the 
negative grains at densities greater than $\sim 10^{10}\cm^{-3}$. 
At densities greater than $\sim 10^{12}\cm^{-3}$ (which is generally
not reached in our computational domain),
Ohmic diffusivity dominates. The large Ohmic diffusivity $\eta_O$ 
is the reason for the ambipolar diffusivity $\eta_A$ to become 
negative above a density $\sim 10^{12}\cm^{-3}$, since it 
contributes negatively to $\eta_A$ (see eq.\ [2]). 
In the opposite extreme of large grain case (LG), where 
all small grains are removed, the Hall diffusivity is positive, 
dominated by the contribution from metal ions. It exceeds the ambipolar
diffusivity above a density $n_H\sim 10^{12}\cm^{-3}$. In this case,
the level of ionization is high enough that the Ohmic diffusivity 
remains unimportant throughout the plotted density range (up to 
$\sim 10^{15}\cm^{-3}$). Overall, the elimination 
of small grains in the LG case makes the magnetic field better 
coupled to the bulk neutral material compared to the MRN case, 
as expected. 

\section{Disk Suppression by a Moderately Strong Magnetic Field}
\label{suppression}

In this section, we will consider a magnetic field of moderate 
strength $B_0=35.4\muG$, corresponding to a dimensionless 
mass-to-flux ratio of $\lambda=2.92$ for the dense ($n_{H_2}=
10^5\cm^{-3}$) core as a whole and $4.38$ for the central 
flux tube. As we argued in \S{}\ref{intro}, this field strength 
is likely on the conservative side. Since the ambipolar diffusivity 
dominates over the Ohmic and Hall diffusivities in most of the density 
range encountered in our calculations (see Fig.\ \ref{Diffusivities}), 
we will first concentrate on its effect on core collapse and disk 
formation in \S{}\ref{PureAD} and \S{}\ref{ADvariation}. Similar AD-only
calculations were performed by \citet{MellonLi2009}, except that we 
now include the prestellar phase of core evolution leading up to 
the formation of a central object (in addition to the protostellar 
phase that they studied) and a more detailed calculation of the 
charge densities, including charged grains. The additional effects of the Ohmic 
and Hall diffusivities are considered in \S{}\ref{Ohmic} and 
\S{}\ref{Hall}, respectively.

\subsection{Reference Model with Ambipolar Diffusion Only}
\label{PureAD}

\begin{figure}
\epsscale{0.45}
\plotone{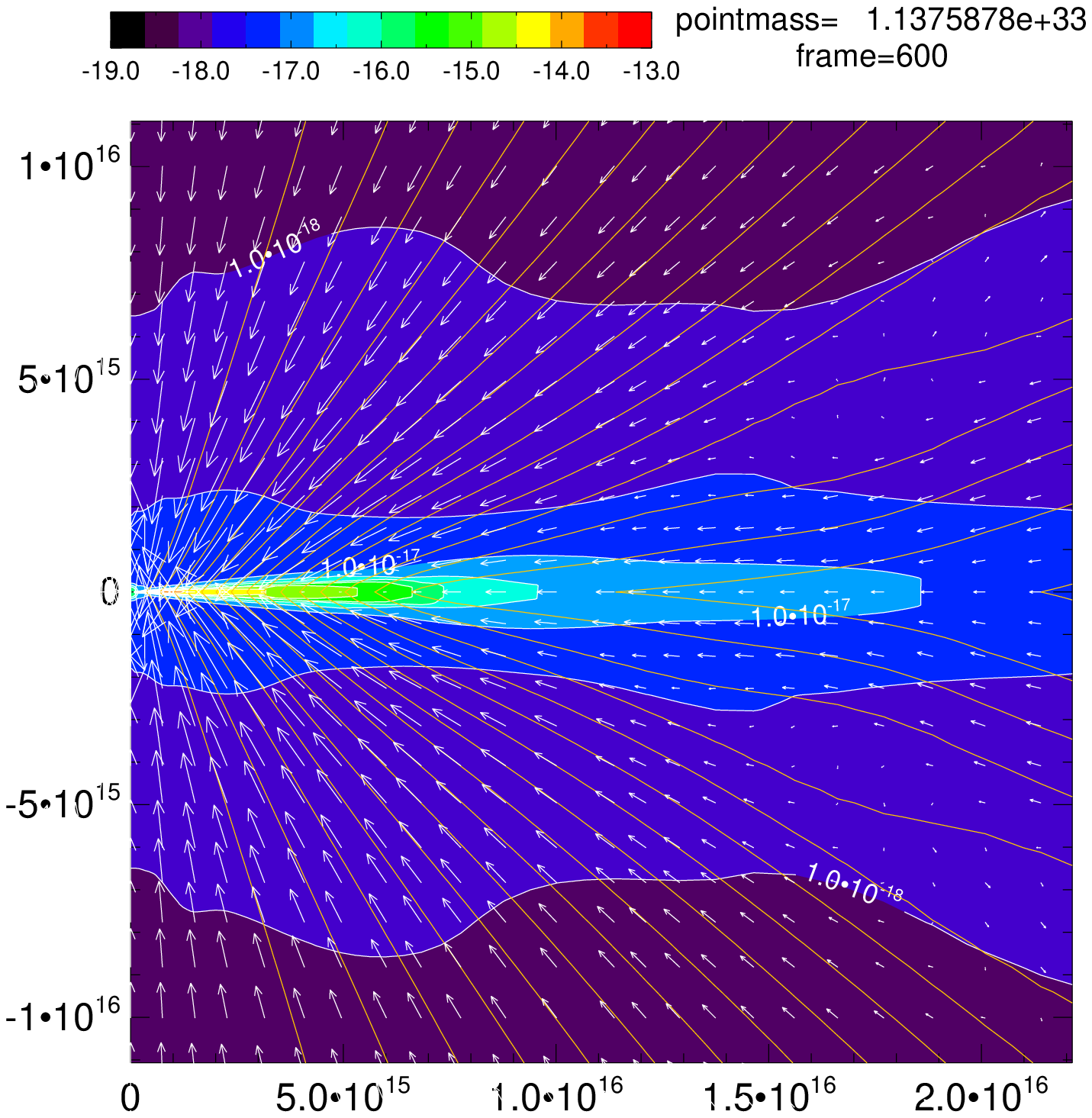}
\plotone{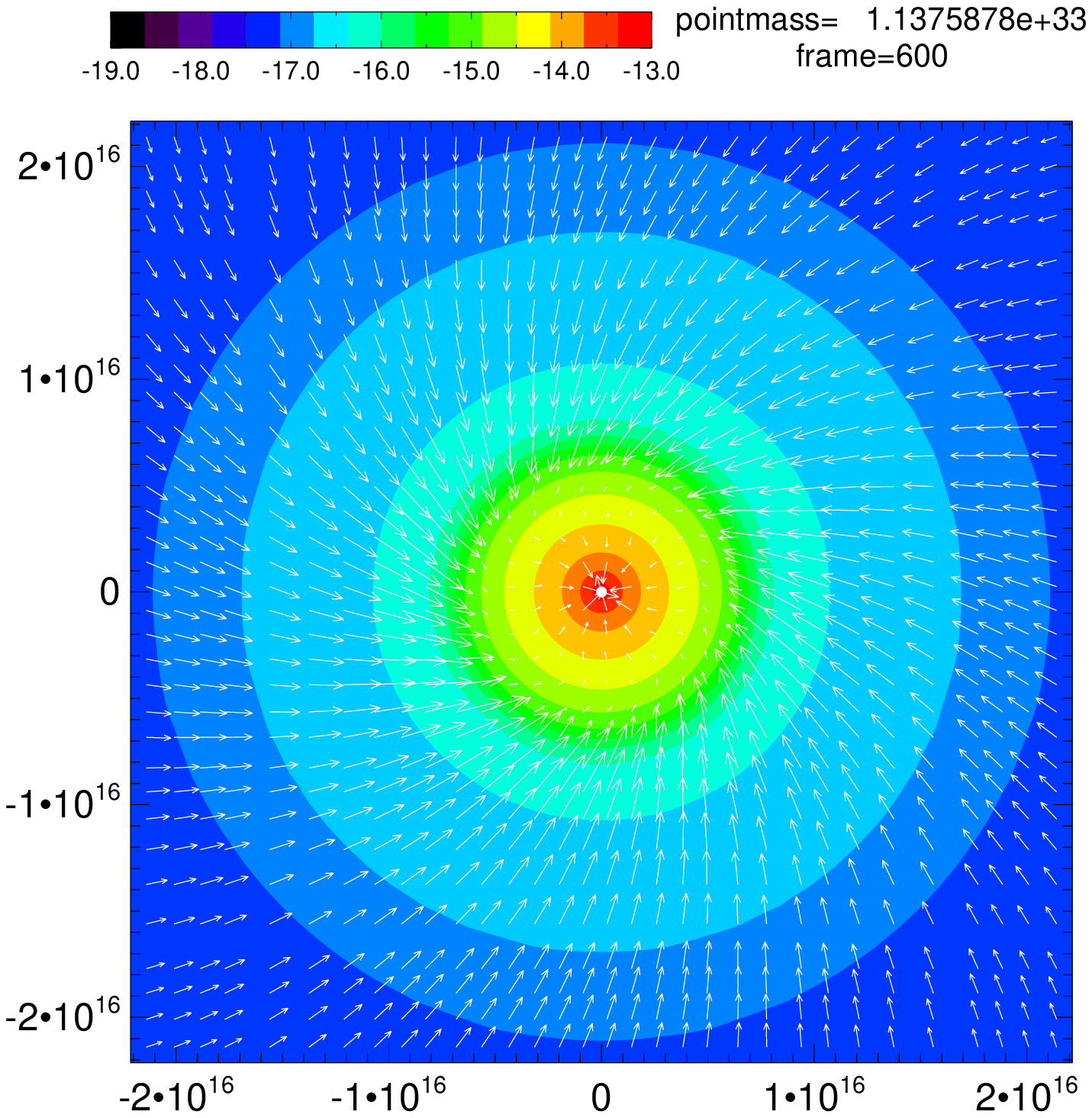}
\caption{Density distribution (color map) and velocity field (white
  arrows) of the reference model (Model REF) in the meridian (left 
panel) and equatorial (right panel) planes, at a representative time 
$t=6\times 10^{12}\second$. The highly flattened, dense equatorial
  structure is not a rotationally supported disk, but rather a 
magnetically supported, nearly non-rotating pseudodisk. Also 
plotted in the left panel are poloidal field lines, with the same 
magnetic flux between adjacent lines. The color bars above the 
panels are for ${\rm log}(\rho)$, with ${\rm{g\,cm}}^{-3}$ and $\rm{cm}$ as the 
units for $\rho$ and length. 
}
\label{ReferenceModel}
\end{figure}

\begin{figure}
\epsscale{0.45}
\plotone{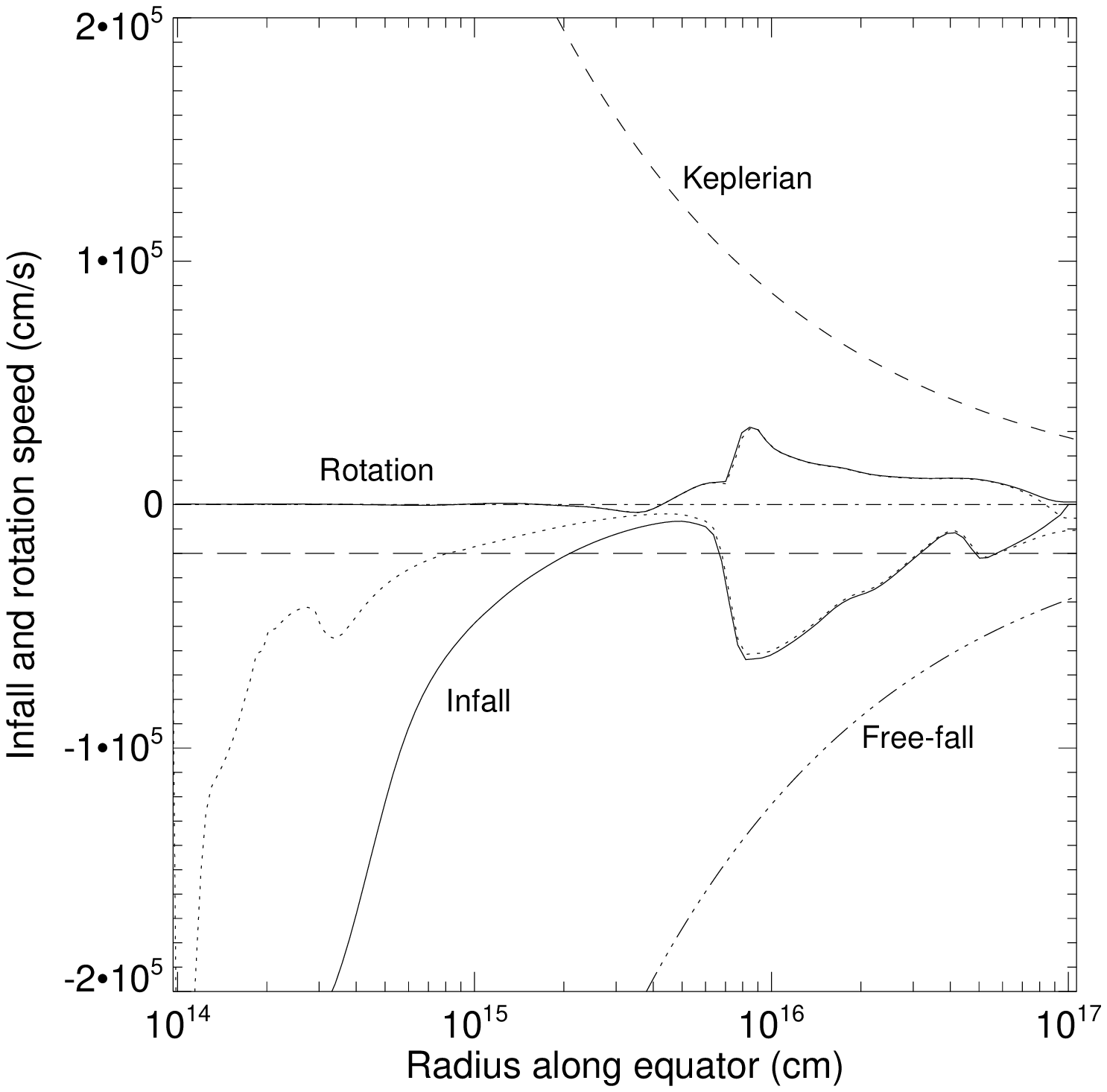}
\plotone{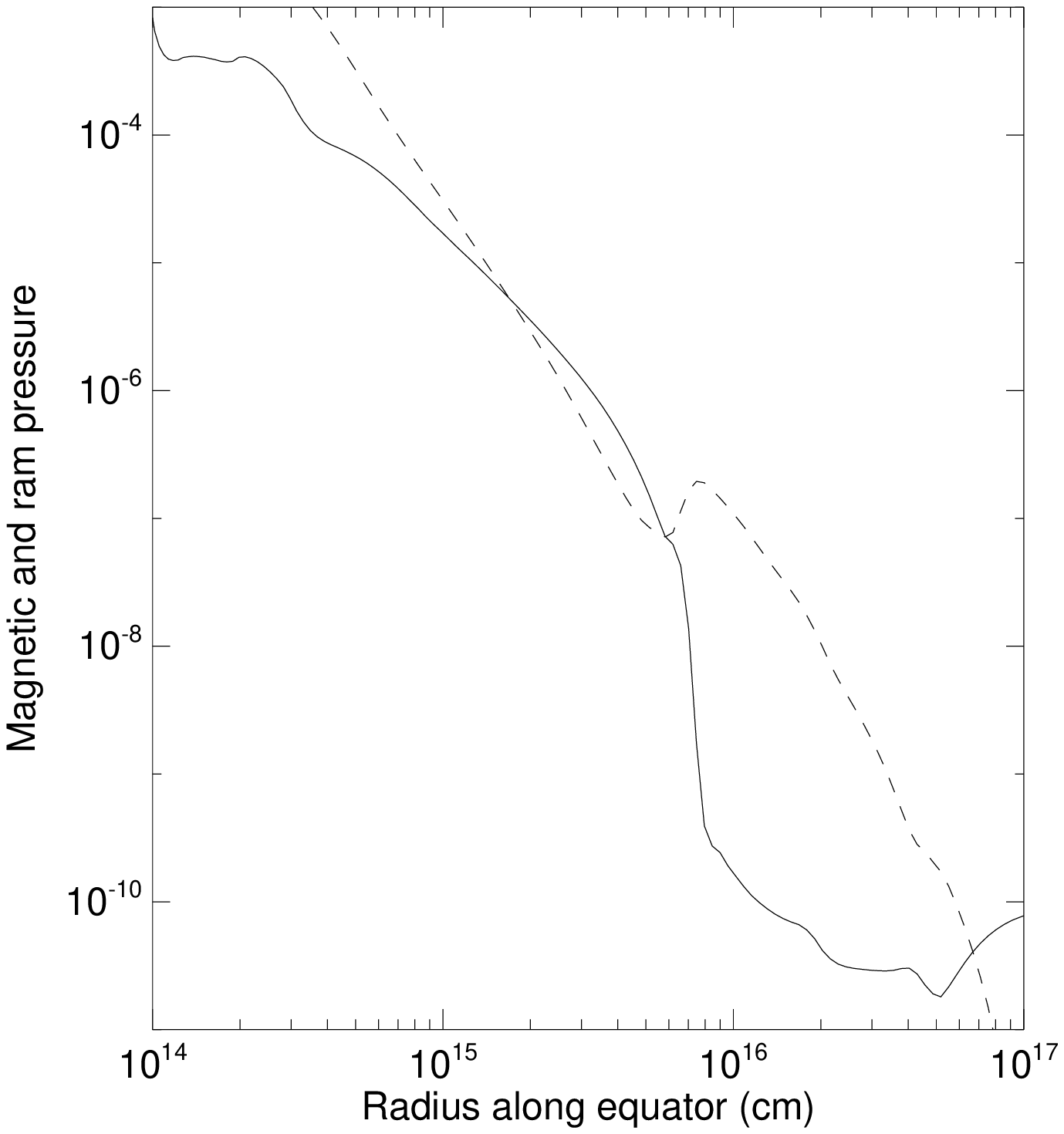}
\caption{Left panel: equatorial infall (lower solid curve) and 
rotation (upper solid) speed for the bulk neutral material in 
the reference model (Model REF) at $t=6\times 10^{12}\second$. Also 
plotted for comparison are the Keplerian 
speed (upper dashed) and free fall speed (lower dash-dotted) 
based on the central mass, the sound speed (horizontal dashed 
line), and zero speed line. The rapid deceleration of infalling 
material near the radius $\sim 8\times 10^{15}\cm$ corresponds 
to a C-shock induced by ambipolar diffusion. The ``effective 
ion speeds'' defined in eq.\ (4) are shown as dotted lines.  
Right panel: magnetic (solid line) and infall ram pressure 
(dashed) along the equator in cgs units, showing a rapid 
increase in field strength near the AD shock and a corresponding 
drop in ram pressure. 
 } 
\label{pressures}
\end{figure}

\begin{figure}
\epsscale{0.8}
\plotone{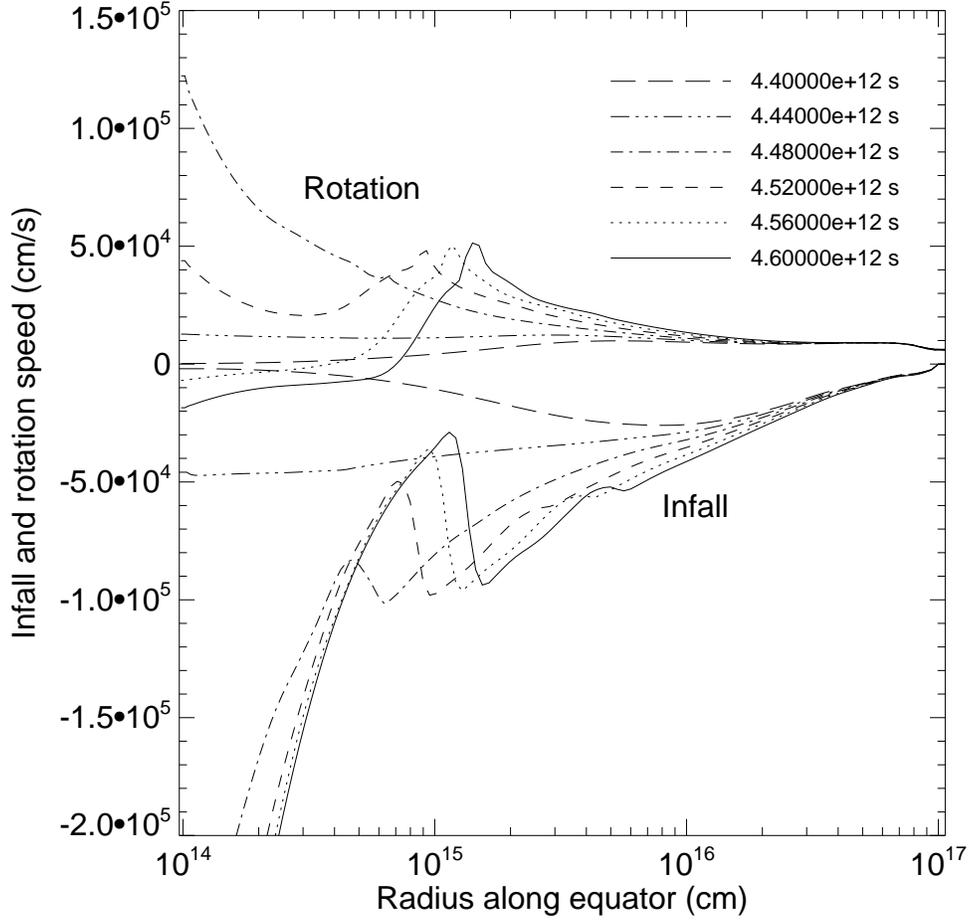}
\caption{Infall and rotation speeds along the equator for the reference model 
(Model REF)  
during the transition from the pre-stellar phase of core evolution 
to the protostellar accretion phase, between $t=4.4\times 10^{12}$ 
and $4.6\times 10^{12}\second$, showing the development of the AD shock
(lower curves) and the associated strong braking of the post-shock 
material (upper curves). 
 } 
\label{transition}
\end{figure}

For the reference model, we adopt the fiducial values for the initial
field strength $B_0=35.4\muG$, cosmic ray ionization rate $\zeta=
10^{-17}\second^{-1}$, and initial core rotation rate $\Omega_0=
10^{-13}\second^{-1}$, as well as an MRN grain size distribution. Since 
grain growth in dense cores tends to make the magnetic field better 
coupled to the neutral matter (see Fig.\ \ref{Diffusivities}) and 
magnetic braking more efficient, our adoption of the MRN distribution 
is also on the conservative side. The parameters for this (Model REF) 
and other models are listed in Table 1. The results are shown in
Figs.\ \ref{ReferenceModel}, \ref{pressures} and \ref{transition}. 

\begin{deluxetable}{llllll}
\tablecolumns{6}
\tablecaption{Model Parameters \label{table:first}}
\tablehead{
\colhead{Model$^{a}$}     & \colhead{Grain}  & \colhead{ $B_0$ ($\mu$G) } & \colhead{
  $\zeta (10^{-17}\second^{-1})$ } & \colhead{$\Omega_0 (10^{-13}\second^{-1})$ }  &
RSD?$^b$
}
\startdata
REF   & MRN  & 35.4  & 1  &  1 & no  \\
LG   & LG  & 35.4  & 1  &  1 & no   \\
LoCR   & MRN  & 35.4  & 0.1  &  1 &  no  \\
HiCR   & MRN  & 35.4  & 3  &  1 &  no  \\
LoROT   & MRN  & 35.4  & 1  &  0.5 & no   \\
HiROT   & MRN  & 35.4  & 1  &  2 & no  \\
REF$_{AO}$   & MRN  & 35.4  & 1  &  1 & no   \\
REF$_{O}$   & MRN  & 35.4  & 1  &  1 & no  \\
REF$_{AHO}^-$   & MRN  & 35.4  & 1  &  1 & no   \\
REF$_{AHO}^+$   & MRN  & 35.4  & 1  &  1 & no \\
WREF   & MRN  & 10.6  & 1  &  1 & no \\
WLG   & LG  & 10.6  & 1  &  1 & no  \\
WHiCR   & MRN  & 10.6  & 3  &  1 &  no  \\
WLoROT   & MRN  & 10.6  & 1  &  0.5 & no   \\
WLoCR   & MRN  & 10.6  & 0.5  &  1 &  yes? \\
WHiROT   & MRN  & 10.6  & 1  &  2 & yes?  \\
VWREF   & MRN  & 3.54  & 1  &  1 & yes? \\
NoROT$_{AHO}^-$   & MRN  & 35.4  & 1  &  0 & no \\
NoROT$_{AHO}^+$   & MRN  & 35.4  & 1  &  0 & no \\
\enddata
\tablecomments{a) The subscript ``A,'' ``H'' and ``O'' denote, 
respectively, ``ambipolar diffusion,'' ``Hall effect,'' and ``Ohmic
  dissipation'' included in a model; models without a subscript 
  include only ambipolar diffusion. The superscript ``-'' 
  (``+'') denotes an initial magnetic field anti-parallel 
  (parallel) to the rotation axis. b) RSD stands for ``rotationally
  supported disk.'' In Model WREF, a small RSD is formed temporarily
  at early times; it is subsequently suppressed by magnetic 
braking. For Models WLoCR, WHiROT and VWREF, an RSD forms early but 
whether it would survive to later times remains unclear because 
of numerical difficulty. }
\end{deluxetable}

Fig.\ \ref{ReferenceModel} displays the density distribution and 
velocity field on the meridian and equatorial planes for the inner 
part ($10^3$-AU scale) of the computation domain, at a representative 
time $t=6\times 10^{12}\second$ (or twice the initial free-fall time), 
when $0.57\Msun$ (or $57\%$ of the 
initial core mass) has fallen to the center. From the left panel, it is 
clear that the density distribution on the meridian plane is highly 
flattened, especially at high densities. The dense, flattened, 
equatorial structure is {\it not} a rotationally supported disk, 
however. Direct evidence against such a disk comes from the right 
panel, which shows a transition from an outer region of rapid 
rotating-infall to an inner region that is neither collapsing nor 
rotating rapidly. The transition is shown more quantitatively in 
the left panel of Fig.\ \ref{pressures}, where the infall and 
rotation speeds on the equator are plotted. The equatorial infall 
is initially slowed down near a relatively large radius 
$r=5\times 10^{16}\cm$. It corresponds to the edge of the magnetic 
bubble inflated by magnetic braking (not shown in
Fig.\ \ref{ReferenceModel}), where a magnetic barrier forces the 
collapsing material over a large solid angle into 
a narrow equatorial channel (see Fig.\ 2 of \ct{MellonLi2008} and 
associated discussion). Upon passing through the barrier, the material
resumes rapid radial infall, spinning up as it collapses, until
a second barrier is encountered at $r\sim 8\times 10^{15}\cm$. 
This second barrier is induced by ambipolar diffusion, which enables 
the magnetic field lines to decouple from matter and pile up outside 
the central object as the matter accretes onto the protostar (see 
discussion in \S{}\ref{intro}). The piled-up magnetic field drives 
a C-shock, which slows down the accretion flow to a subsonic speed. 
The slowdown of infall leads to a high density in the postshock 
region, which is clearly visible in the density maps 
(Fig.\ \ref{ReferenceModel}). Upon passing through the shock, the 
equatorial material re-accelerates towards the central object, 
reaching a highly supersonic infall speed at small radii. The 
supersonic infall clearly indicates that a rotationally supported 
disk (RSD hereafter for short) is not present.   

The lack of an RSD is even more obvious from the rotation speed 
on the equator plotted in Fig.\ \ref{pressures}. As the infalling 
material enters the AD shock, its 
rotation speed drops quickly. Over most of the postshock region, the 
rotation speed is nearly zero, indicating an efficient braking of 
the material that accretes onto the central object through the 
equatorial region. 

Why is it that the rotation of the infalling material is braked 
almost completely? The answer lies in the strong magnetic field 
trapped interior to the AD shock (\ct{LiMcKee1996};
\ct{Contopoulos_1998}; \ct{KrasnopolskyKonigl2002}).   
The field trapping is already evident in the left panel of 
Fig.\ \ref{ReferenceModel}, which shows a pileup of poloidal 
magnetic field lines at small radii. It is shown more quantitatively in 
the right panel of Fig.\ \ref{pressures}, where the distribution of 
the magnetic pressure ($B^2/[8\pi]$) with radius is plotted along 
the equator. Note the rapid increase  
in magnetic pressure (and thus field strength) across the AD shock 
(near $r_s\sim 8\times 10^{15}\cm$). There is a corresponding drop 
in the ram pressure ($\rho v_r^2$), indicating that 
the strong post-shock 
magnetic field is trapped by the ram pressure of the pre-shock 
collapsing flow \citep{LiMcKee1996}. The further rise in the ram 
pressure at smaller radii is due to the gravity of the central 
object, which re-accelerates the equatorial material (that was 
temporarily slowed down to a subsonic speed behind the AD shock) 
to a highly supersonic speed. The ``effective ion speed'' in the 
post-shock region is much lower than that of the bulk neutral 
material (see the right panel of Fig.\ \ref{ReferenceModel}). The 
relatively large drift velocity $\Delta {\bf v}_{AD}$, driven by 
a large outward magnetic force, is what enables the magnetic flux to
accumulate in the postshock region in the first place: as more and 
more matter accretes onto the central object, more and more magnetic 
flux is left 
behind. Indeed, by the time shown in Figs.\ \ref{ReferenceModel} 
and \ref{pressures}, about half of the total magnetic flux of the 
initial core is confined within the shock radius $r_s\sim 8\times 
10^{15}\cm$. The postshock region, which contains a small fraction 
($\sim 4\%$) of the total mass, is so strongly magnetized that 
it becomes highly magnetically subcritical (with a local 
dimensionless mass-to-flux ratio $\lambda$ well below unity) 
even though the whole core was significantly magnetically supercritical
to begin with; the small infall speed (much below free fall value) 
and nearly vanishing rotation speed indicate that the material in
the region is essentially magnetically supported. 
The strong postshock field is only part of the reason for efficient
braking. Another part is that the postshock field naturally bends 
outwards due to strong equatorial pinching (see the left panel of
Fig.\ \ref{ReferenceModel}), which increases its lever arm and thus the
braking efficiency.

Even though the equatorial rotation speed at small radii remains close to
zero at late times (as shown in Figs.\ \ref{ReferenceModel} and 
\ref{pressures}), it does reach a substantial, 
supersonic, value for a brief period of time during the transition
from the phase of prestellar collapse to the protostellar accretion 
phase. Fig.\ \ref{transition} 
displays the equatorial infall and rotation speeds during the 
transition, for 6
times separated by $\Delta t= 4\times 10^{10}\second$. At the earliest of
the times shown, the material at small radii remains static, as
expected during the pre-stellar phase of core evolution for the 
region within one thermal Jeans length of the origin.  When the
central Jeans 
length shrinks to inside our inner boundary, rapid protostellar accretion
ensures, leading to a quick spin-up of the collapsing material. The
rotation speed reaches a value as high as $\sim 1 \kms$ (or $\sim
5$ times the sound speed) near the inner boundary before decreasing 
back again. The rapid, transient spin-up is a feature that is not 
captured by the self-similar solutions of \citet{KrasnopolskyKonigl2002}
and \citet{MellonLi2009}. The spindown is clearly associated 
with the development of an ambipolar diffusion induced accretion 
shock (see the curves for infall
speed), which strengthens as it propagates outward. The material in 
the post-shock region is so strongly braked that it rotates backwards
for a short period of time, before settling down to the nearly 
non-rotating state shown in Figs.\ \ref{ReferenceModel} and 
\ref{pressures}. The strong magnetic braking prevents any rotationally 
supported disk larger than $10^{14}\cm$ (the size of our inner 
boundary) from forming at any time during our (long) simulation, which 
lasted until $t=9\times 10^{12}\second$, when $90\%$ of the core mass has 
been accreted by the central object. 

\subsection{Pure AD Runs: Grain Size, Cosmic Ray Ionization and Rotation Rates}
\label{ADvariation}

In the reference model, the magnetic braking has clearly suppressed
the formation of a rotationally supported disk (although a highly 
flattened, dense, magnetically supported, nearly non-rotating 
pseudodisk was formed). In this subsection, 
we explore how robust this result is, by varying the physical 
quantities that may affect disk formation in a lightly ionized, 
rotating, magnetized molecular cloud core. These include the grain 
size distribution and cosmic ray ionization rate, both of which 
affect the charge densities and thus the degree of coupling between 
the magnetic field and the bulk neutral matter, as well as the rate 
of core rotation. We discuss models where these quantities are 
varied over a reasonable range (see Table 1).  

As discussed in \S{}\ref{s:charge}, the size distribution of the grains 
in dense cores of molecular clouds is uncertain. We have adopted in 
our reference model the standard power-law MRN distribution that 
includes a large amount of small grains. In this subsection, we will 
consider an opposite limit, Model LG in Table 1, where all grains are
assumed to be $1\mum$ in size and the small grains are completely 
absent. Besides grain size, the charge densities are also affected 
by the cosmic ray ionization rate $\zeta$. \citet{Padovani_2009} 
compiled from the literature the values of $\zeta$ inferred for clouds 
of a wide range of column densities, from diffuse clouds to massive 
protostellar envelopes (see their Fig.\ 15). Most of the inferred 
values are above our reference value $\zeta=10^{-17}\second^{-1}$, 
although there are a few exceptions. We will consider a model with 
$\zeta=10^{-18}\second^{-1}$ (Model LoCR in Table 1), which is 
a lower limit to the inferred values. For illustration, 
we will also consider a case with a higher rate $\zeta=3\times 
10^{-17}\second^{-1}$ (Model HiCR), close to the value inferred by 
\citet{Webber1998} for the local interstellar medium. In addition, we 
consider the effect of varying the core rotation rate
$\Omega_0$. Our reference value $\Omega_0=10^{-13}\second^{-1}$ 
corresponds to a ratio of rotational to gravitational binding energy 
of $\beta=0.025$, which is typical of the cores discussed in
\citet{Goodman_1993}.
A spread exists for the inferred 
$\beta$ values, which motivates us to consider two additional  
rotation rates: $\Omega_0=5\times 
10^{-14}$ (Model LoROT) and $2\times 10^{-13}\second^{-1}$ (Model 
HiROT), corresponding to $\beta=0.006$ and $0.1$, respectively. 

\begin{figure}
\epsscale{0.8}
\plotone{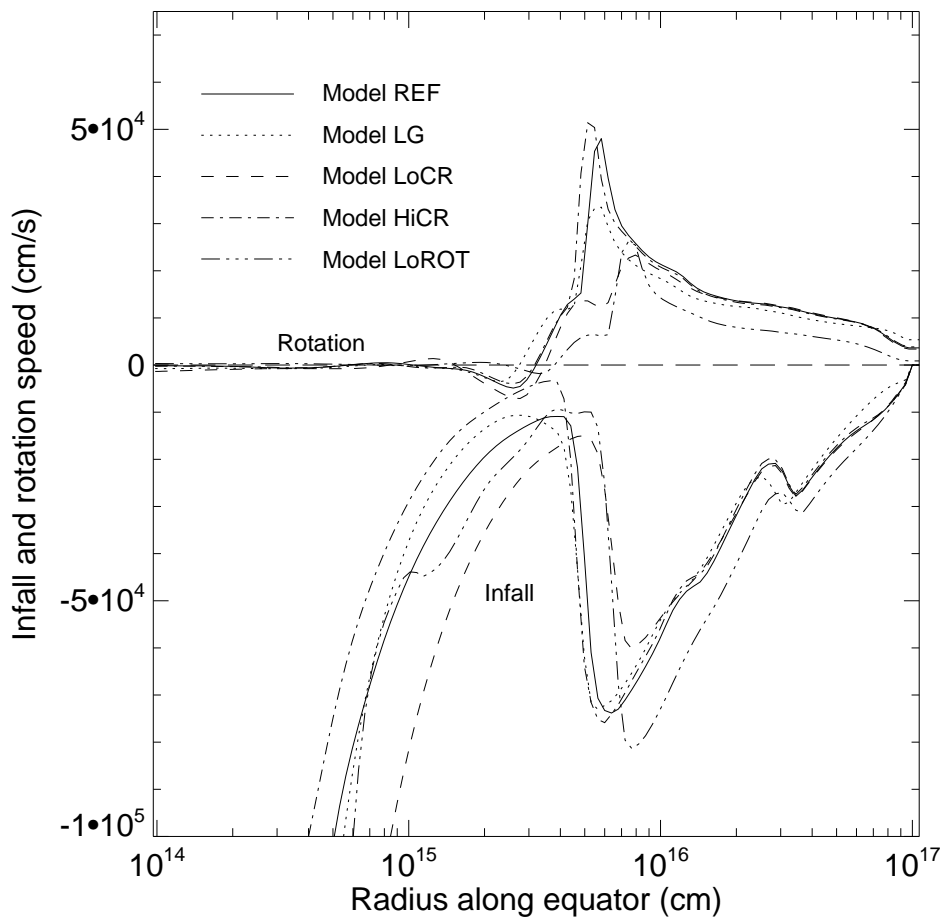}
\caption{Comparison of the infall and rotation speeds along the 
equator for variants of the reference model with different 
grain size distribution (Model LG), cosmic ray ionization
rate (Models LoCR and HiCR), and rotation rate (Model 
LoROT), at a representative time $t=5.5\times 10^{12}\second$. 
The nearly zero rotating speed at small radii indicates the 
absence of a rotationally supported disk due to efficient 
magnetic braking for all cases shown.  
 } 
\label{general}
\end{figure}

The different variants of the reference model are compared in 
Fig.\ \ref{general}, which plots
the infall and rotation speeds on the equator at a common time
$t=5.5\times 10^{12}\second$ for all models except Model HiROT (which has 
yet to form a central object at this time, because the collapse is 
significantly delayed by the combination of a moderately strong 
magnetic field and fast rotation).  
The most striking feature of Fig.\ \ref{general} is that the rotation
speed is essentially zero at small radii (within $\sim 10^{15}\cm$) 
for all of the models shown. The same is true for Model HiROT (not 
shown in the figure) at later times (greater than $\sim 6.2\times 10^{12}\second^{-1}$), 
when the fast rotating core has collapsed. Apparently the magnetic 
braking is strong enough to remove essentially all of the angular 
momentum of the equatorial material inside the AD shock for the 
realistic ranges of grain size distribution, cosmic ray ionization 
rate and core rotation rate explored here. 

Suppression of RSD in some cases is easy to understand. The
elimination of small grains in Model LG and the higher cosmic 
ray ionization rate in Model HiCR increase the densities of 
electrons and ions, which strengthens the coupling between 
the magnetic field and the bulk neutral matter. Since a better 
magnetic coupling is expected to make the braking more 
efficient, it is not surprising to find that the RSD remains 
suppressed as in the reference model. In these two cases, the 
AD shock is located at a slightly smaller radius compared with 
the reference case (see Fig.\ \ref{general}). We interpret 
this effect as a result of the better coupling, which 
forces the bulk neutral material to collapse more slowly, 
which in turn leads to a 
somewhat lower central mass ($8.62\times 10^{32}\gram$ for 
Model LG and $8.69\times 10^{32}\gram$ for Model HiCR) than 
in the reference model at the same time ($9.29\times 10^{32}\gram$). 
Associated with the lower mass is a lower magnetic flux 
accumulating outside the central object, which drives a smaller 
AD shock. 

The situation with Model LoCR is the opposite. The lower cosmic 
ray ionization rate decreases the degree of ionization in the 
core, which weakens the coupling between the magnetic field 
and the bulk neutral matter. The weakening of magnetic coupling 
is expected to reduce the braking efficiency in principle. The 
braking is not weakened enough, however, to enable an RSD 
to form, as evidenced by the vanishing rotation speed and 
fast infall inside the AD shock. In this case, the shock radius 
is somewhat larger than that of the reference model (see 
Fig.\ \ref{general}), because of a faster post-shock infall, 
which leads to a larger central mass ($9.76\times 10^{32}\gram$ 
for Model LoCR) and thus a larger left-behind magnetic flux 
for shock driving.

Neither Model LoROT nor HiROT produced a rotationally 
supported disk.
The former is to be expected, since the angular momentum is lower 
for a more slowly rotating core, and should be easier to remove by 
magnetic braking. The
slower rotation also presents a lesser obstacle to the collapse, 
allowing the formation of a larger central mass ($1.24 
\times 10^{33}\gram$) and a larger AD shock (see Fig.\ \ref{general}) 
than in the reference 
model at the same time. In the opposite case of HiROT, the stronger
rotational support delayed the formation of a central object 
until a much later time. Once the central object has formed,  
magnetic braking is again strong enough to remove essentially all 
of the angular momentum from the material near the central object, 
preventing an RSD from forming during the protostellar accretion 
phase. 

We conclude that disk formation is suppressed by a moderately strong 
magnetic field in the presence of 
ambipolar diffusion for the range of parameters that we consider 
realistic. 

\subsection{Ohmic Dissipation} 
\label{Ohmic}

\begin{figure}
\epsscale{0.8}
\plotone{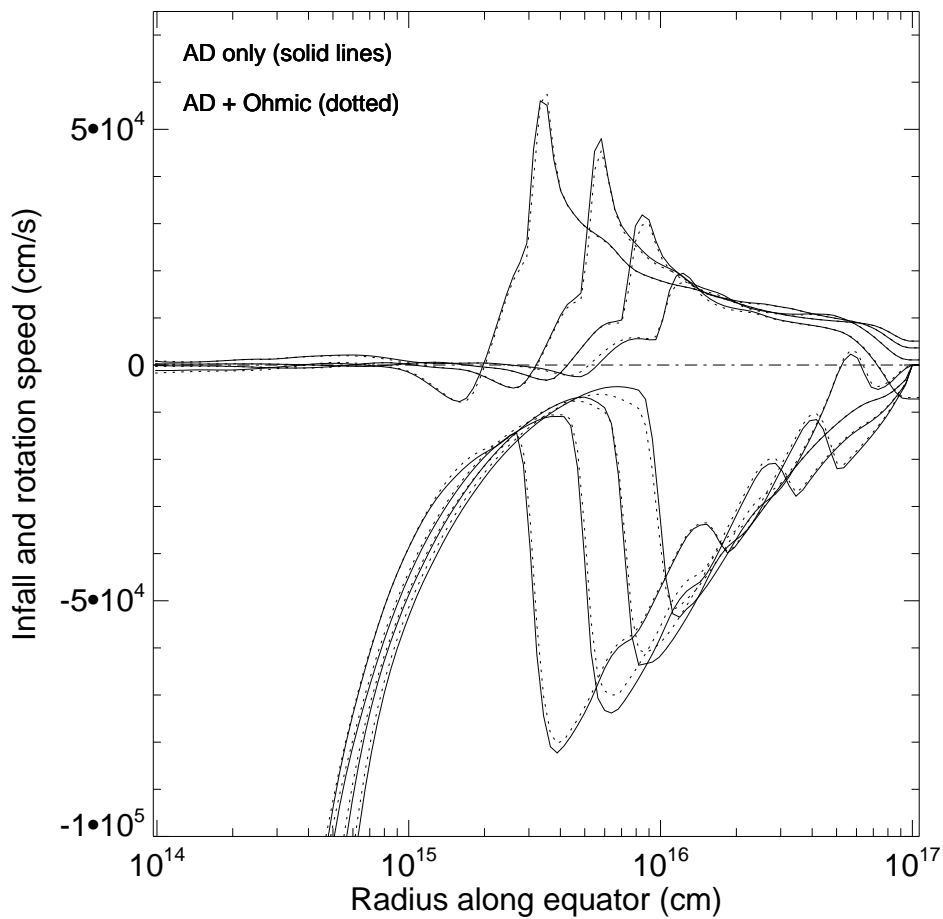}
\caption{Comparison of the infall and rotation speeds along the
 equator for Model REF$_{AO}$ (with both ambipolar diffusion and 
Ohmic dissipation; dotted lines) to the reference model (with only
ambipolar diffusion; solid), at representative times $t=5\times
 10^{12}$, $5.5\times 10^{12}$, $6\times 10^{12}$, and $6.5\times 
10^{12}\second$. The similarity between the two sets of curves shows 
that Ohmic dissipation does not affect the flow dynamics much. 
 } 
\label{AO}
\end{figure}

How does the Ohmic term in the induction equation (1) affect the core
collapse and disk formation? To address this question, we repeat  
the reference model, but with the Ohmic dissipation included (Model 
REF$_{AO}$ in Table 1) in addition to the ambipolar diffusion 
(although not yet the Hall term, see below). It turns out that the Ohmic 
dissipation modifies the dynamics of the core collapse and
protostellar accretion relatively little, as can be seen from Fig.\ \ref{AO},
which compares the equatorial infall and rotation speeds of Model REF$_{AO}$ 
and the reference model at four representative times. The speeds for
the two cases are barely distinguishable over most of the space. 
The conclusion is that in the presence of ambipolar diffusion, Ohmic 
dissipation does not enable the formation of RSD. 

\begin{figure}
\epsscale{0.8}
\plotone{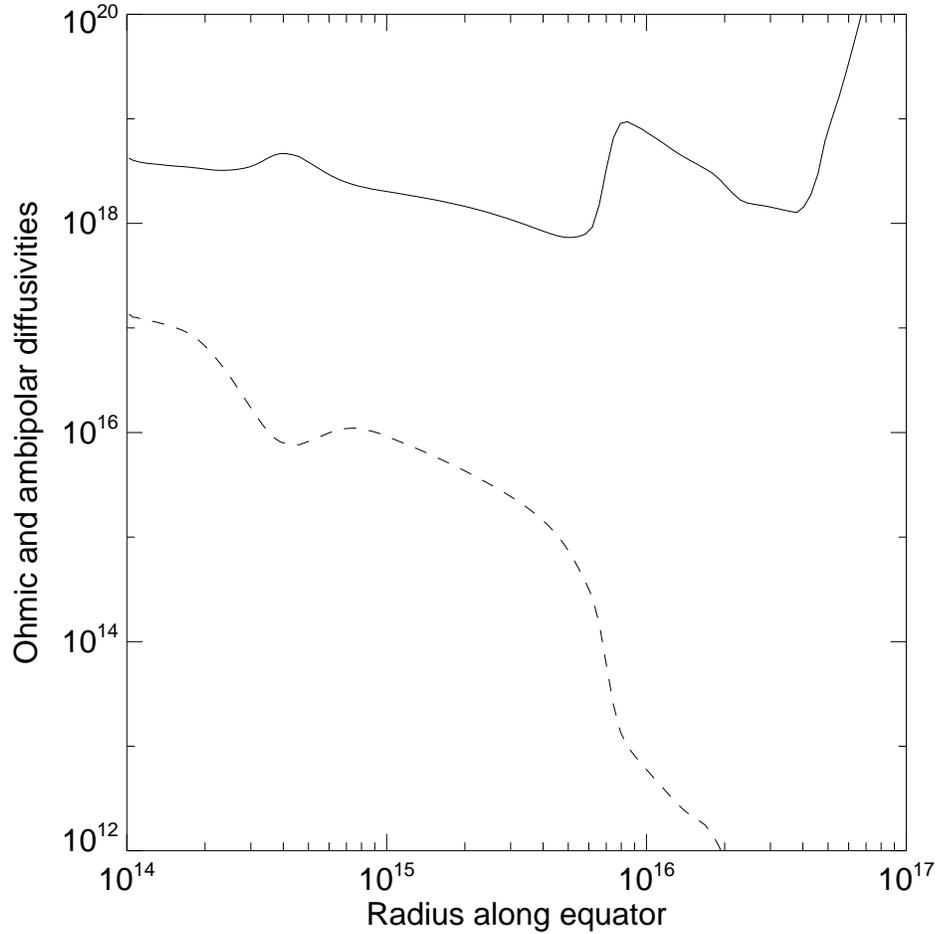}
\caption{Comparison of the Ohmic diffusivity (dashed line) and the 
ambipolar diffusivity (solid) along the
 equator for Model REF$_{AO}$ (with both ambipolar diffusion and Ohmic dissipation) at a representative time $6\times 10^{12}\second$. The large difference between 
the two indicates that the Ohmic dissipation is relatively unimportant. 
 } 
\label{AOvalues}
\end{figure}

The reason that Ohmic dissipation is ineffective in modifying the 
flow dynamics in the presence of ambipolar diffusion is simple: the 
Ohmic diffusivity $\eta_O$ is smaller than the ambipolar diffusivity
$\eta_A$ by more than an order of magnitude (see Fig.\ \ref{AOvalues}). 
The relatively small Ohmic diffusivity is a result of the relatively 
low density in the computational domain, with  $\rho 
\lesssim 10^{-13}\gram\cm^{-3}$ (or $n_H \lesssim 4\times 
10^{10}\cm^{-3}$) typically. 
The moderate density is a result of strong magnetic braking, which 
yields a relatively low-density, collapsing pseudodisk (rather than 
a denser, well supported  RSD) in the equatorial region.  

\begin{figure}
\epsscale{1.2}
\plottwo{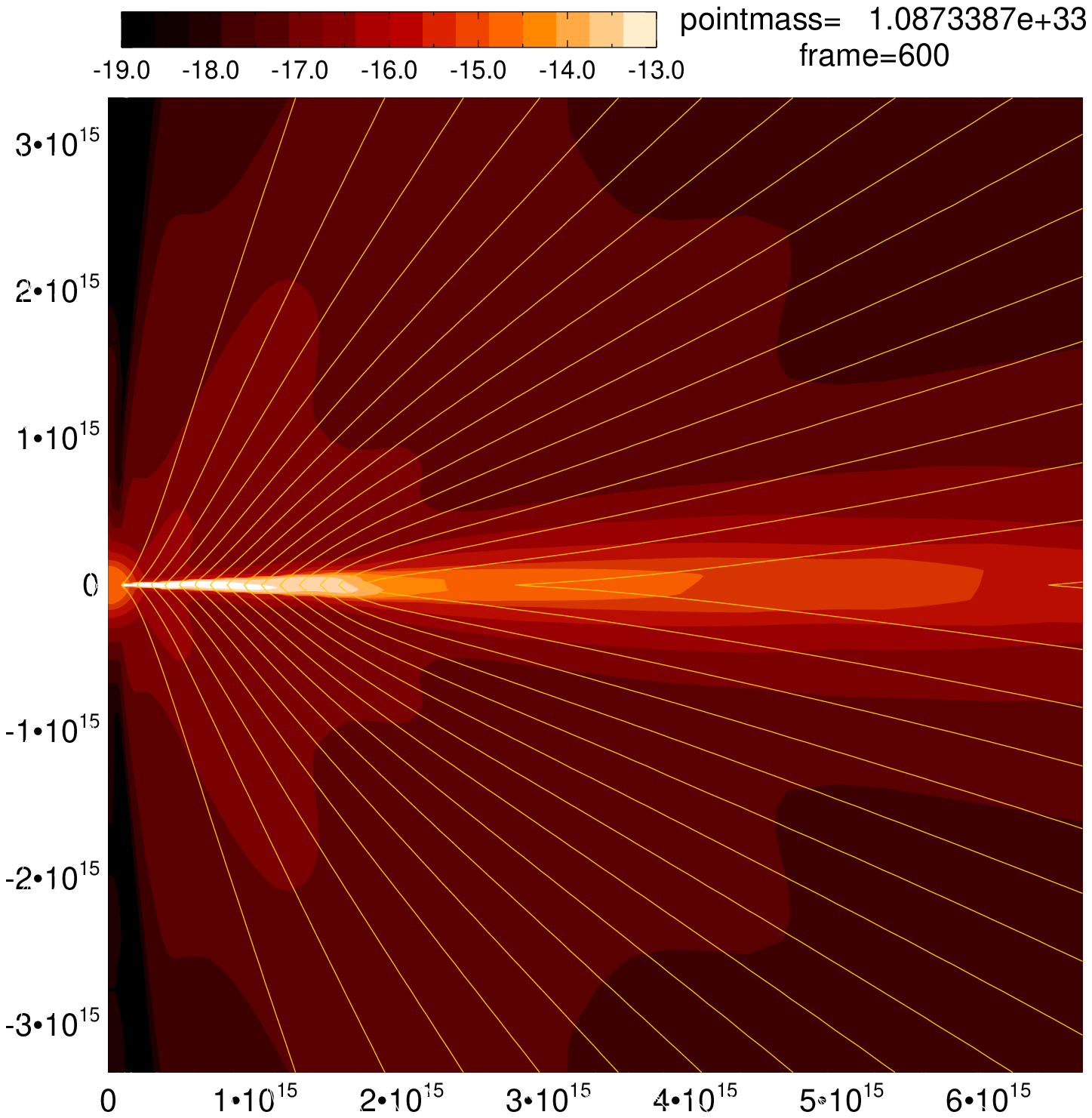}{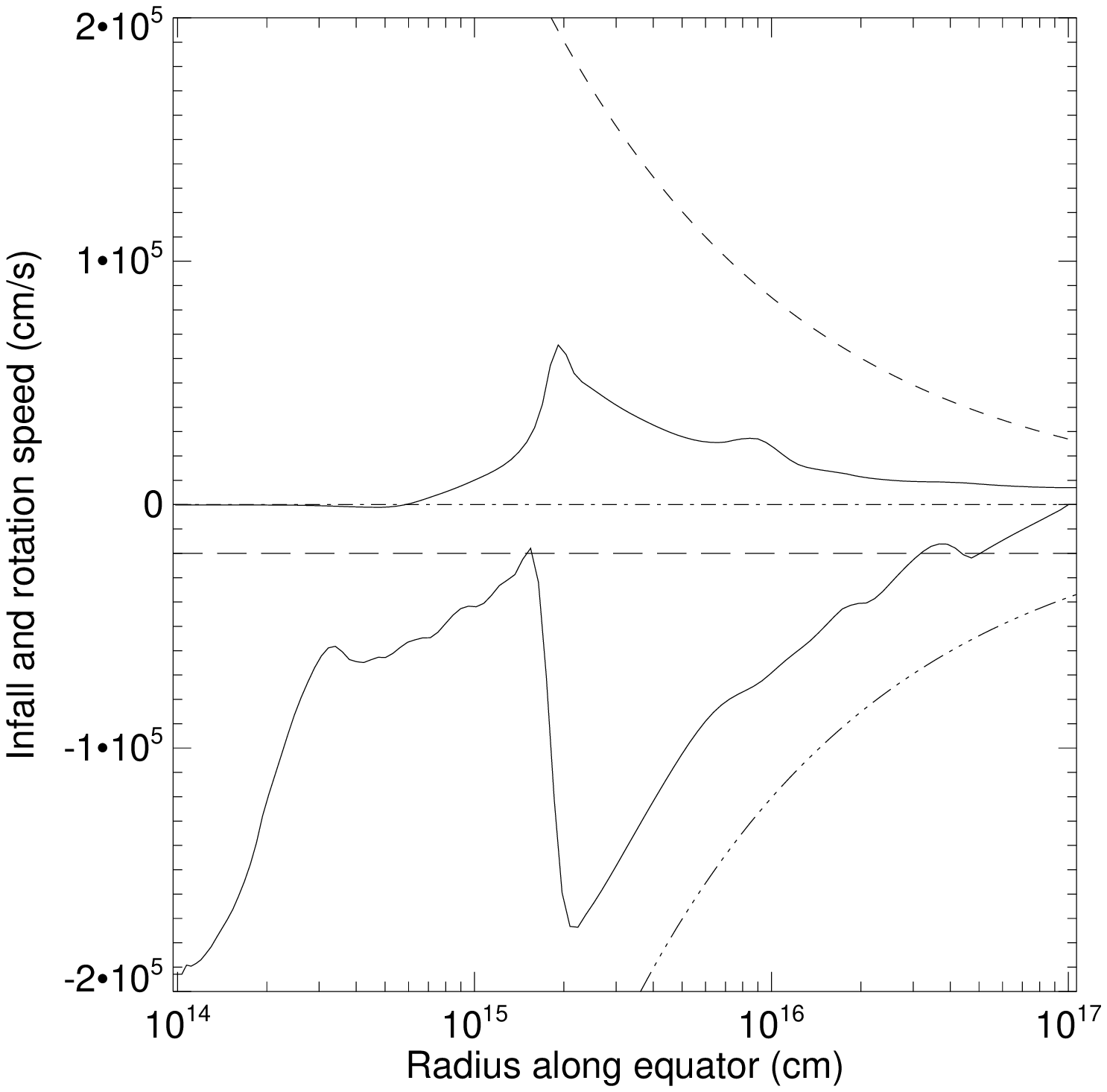}
\caption{Density map with field lines (left panel) and equatorial infall and
  rotation speed (right) for Model REF$_O$ (with Ohmic dissipation
  only) at $t=6\times 10^{12}\second$.  
Also plotted for comparison in the right panel are the Keplerian 
speed based on the 
central mass (upper dashed curve), free fall speed (lower dash-dotted 
curve), sound speed (horizontal dashed line), and zero speed line.    
Note the rapid deceleration of infalling material near the radius $\sim
  2\times 10^{15}\cm$. It is caused by magnetic flux pileup at
  small radii due to Ohmic diffusion, similar to the AD shock shown 
in Figs.\ \ref{ReferenceModel} and \ref{pressures}. }
\label{PureO}
\end{figure}

Even when ambipolar diffusion is turned off, we do not find any 
rotationally supported disk within our computation domain 
(outside a radius of $10^{14}\cm$) that is resulted from Ohmic
dissipation (Model REF$_O$ in Table 1). Indeed, the overall 
flow dynamics in the pure Ohmic case is strikingly similar to 
the reference case that has only ambipolar diffusion (compare 
Fig.\ \ref{PureO} to Fig.\ \ref{ReferenceModel}). The reason for 
the similarity is that Ohmic resistivity 
does not destroy the {\it net} magnetic flux that passes from one 
hemisphere to the other through the equatorial plane (because 
the poloidal field lines are tied to the low density regions well 
above and below the equatorial plane where Ohmic dissipation is
negligibly small; see 
discussion in \ct{Shu_2006}). Rather, it enables the
poloidal field lines to diffuse radially outward, similar to ambipolar
diffusion. As more and more matter accretes across the field lines 
into the center, the left-behind magnetic flux piles up at small 
radii, as shown in  the left panel of Fig.\ \ref{PureO}. The flux 
pile-up is qualitatively similar to the pure-AD reference case (see 
the left panel of Fig.\ \ref{ReferenceModel}). It leads to a strongly
magnetized equatorial region where the infall speed decreases to well 
below the free-fall value and the rotation is almost completely
braked (see the right panel of Fig.\ \ref{PureO}). We have repeated 
the Ohmic dissipation-only calculation with a reduced radius
for the inner boundary (from $10^{14}\cm$ to $3\times 10^{13}\cm$ or
$2\AU$), and found the same result: namely, a realistic level of
(classical) Ohmic resistivity is not large enough to enable the 
formation of rotationally supported disks larger than several AUs, 
at least in 2D (axisymmetry). This is in agreement with \citet{Krasnopolsky_2010},
who showed that enhanced resistivity is needed 
for such disks to form. 

\subsection{Hall Effect}
\label{Hall}

\begin{figure}
\epsscale{0.8}
\plotone{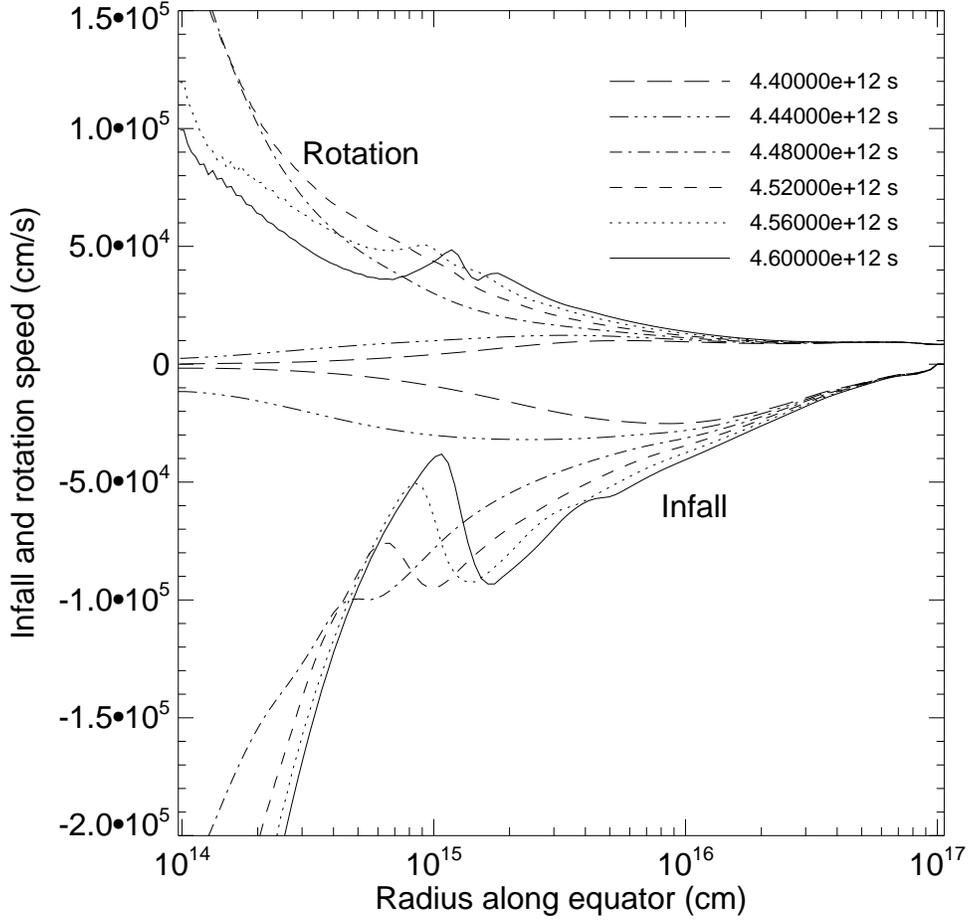}
\caption{Infall and rotation speeds along the equator for Model REF$_{AHO}^-$
 (including all three non-ideal MHD terms) during the transition from 
 the pre-stellar core
 evolution phase to the protostellar accretion phase, between
 $t=4.4\times 10^{12}$ and $4.6\times 10^{12}\second$, as in Fig.\ \ref{transition}. 
 The Hall effect enables the material at small radii inside the AD
 shock to rotate at supersonic values.} 
\label{HallN}
\end{figure}

Under the conditions encountered in our core collapse calculations,
the magnitude of the Hall diffusivity in the induction equation (1) is 
typically larger than the Ohmic diffusivity (see 
Fig.\ \ref{Diffusivities}). We 
therefore expect 
the Hall term to have a larger effect on
the collapse dynamics than the Ohmic term discussed in the last 
subsection. We find that this is indeed the case. Fig.\ \ref{HallN} 
plots the infall and rotation speeds on the equator for Model
REF$_{AHO}^-$, which includes all three nonideal MHD terms 
in equation (1) and an initial magnetic
field that points in a direction opposite to the initial rotation axis
(in the negative ``z'' direction, and hence the
superscript ``-'' in the model name). Plotted are the speeds 
during the transition from the pre-stellar core evolution to 
the protostellar accretion phase, as 
in Fig.\ \ref{transition} for the reference (AD only) model. Comparing 
Figs.\ \ref{HallN} and \ref{transition} reveals that the Hall term 
has relatively little effect on the equatorial infall speed. In both 
cases, as more and more mass accumulates at the center, an AD shock
develops where the collapsing material slows down temporarily, before
reaccelerating towards the origin. The effect on the rotation speed 
is much more pronounced. The Hall effect enabled the equatorial 
material in the post-AD shock region to rotate faster compared to 
the reference case. The difference is especially clear at later times, 
when the rotation inside the AD shock is almost completely braked (or
even reversed) by magnetic braking in the reference case. In the
presence 
of the Hall
effect, the post-shock material rotates at a speed as high as
5 times the sound speed. The rotation speed remains much smaller than
the Keplerian speed, however, indicating that a rotationally supported
disk is not enabled by the Hall effect. The lack of an RSD is also 
evident from the fact that the post-shock region collapses at a high
speed at small radii.  

The significant Hall effect on the post-shock rotation speed can be
understood as follows. The Hall effect depends on the current density 
${\bf j}$ ($\propto \nabla\times {\bf B}$; see eq.\ [1]), which is 
predominantly in the toroidal direction because of strong magnetic 
field pinching in the equatorial region. It is particularly strong 
in the equatorial post-shock region, where the field is 
strong and is bent outward significantly (see the left panel of 
Fig.\ref{ReferenceModel}). The toroidal current drives a twist of 
the field lines in the azimuthal direction, which in turn generates 
a torque that acts to spin up the post-shock material. 

\begin{figure}
\epsscale{0.8}
\plotone{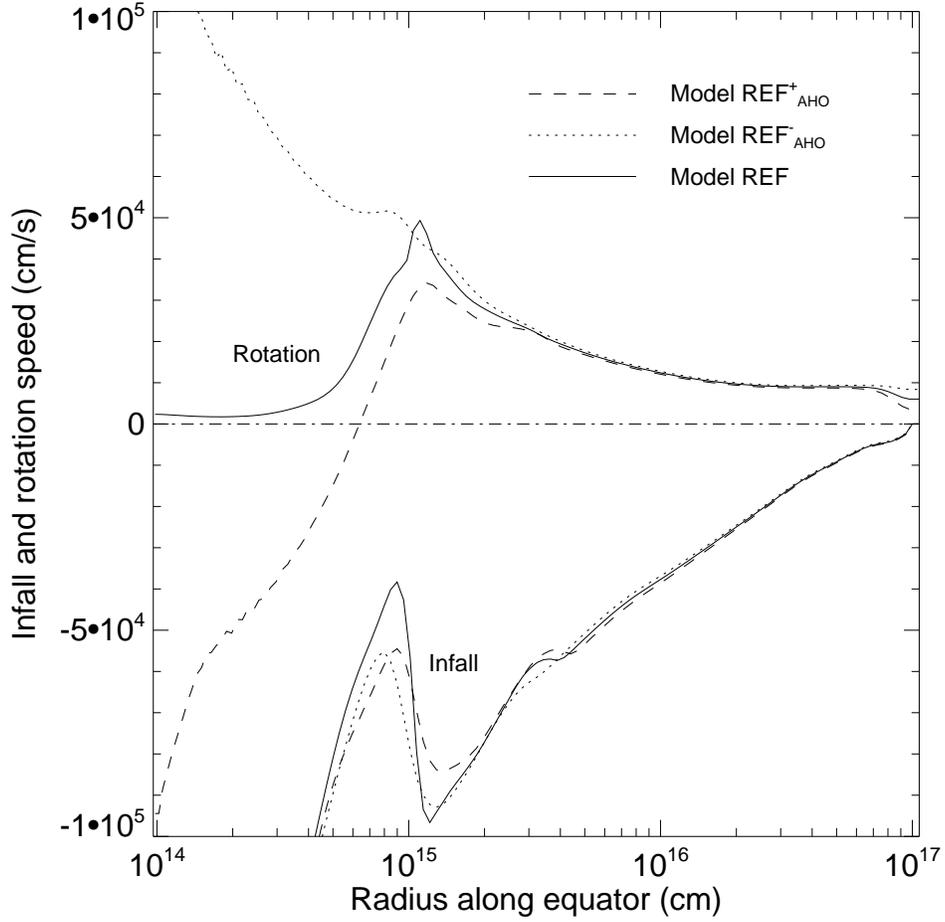}
\caption{Infall and rotation speeds along the equator for two models
 of opposite initial magnetic orientation, Model REF$_{AHO}^+$ (dashed lines)
 and Model REF$_{AHO}^-$ (dotted), at $t=4.55\times 10^{12}\second$. The reference 
model (solid) without the Hall effect is also plotted for comparison. 
 } 
\label{HallP}
\end{figure}

An interesting feature, pointed out for example in the disk-wind study 
of \citet{WardleKonigl1993},
is that when the field direction is reversed, the torque induced by 
the Hall effect changes 
direction as well \citep{Krasnopolsky_2011}. This is
illustrated in Fig.\ \ref{HallP}. Whereas the reference model has a 
post-shock region nearly completely braked in the absence of the Hall 
effect, both Model REF$_{AHO}^+$ and Model REF$_{AHO}^-$ have substantial, 
supersonic
rotation in the post-shock region, although in opposite directions. In 
particular, in Model REF$_{AHO}^+$ where the initial magnetic field 
is aligned (rather than anti-aligned) with the initial rotation axis, 
the magnetic torque induced by the Hall effect has forced the equatorial
material at small radii to rotate in a direction opposite to the
material at larger distances; the resultant shear may induce
instabilities in 3D that should be investigated in the future. 
The change in the direction of the torque 
from Model REF$_{AHO}^-$ to Model REF$_{AHO}^+$ is due to the change 
in the poloidal field direction which, for the 
same outward bending of field lines, produces a flip in the direction
of the toroidal current. Nevertheless, in neither field orientation 
was the Hall spin-up strong enough to produce a rotationally supported 
disk, as evidenced by the rapid infall speed at small radii where the 
rotation speed is relatively high (but still well below the local 
Keplerian speed). Indeed, there is relatively little change in the 
infall speed with or without the Hall effect, because the radial 
current density is relatively small, making the Hall effect less 
important in that direction. 

We will return to the Hall effect in section \ref{PureHall}, where we 
consider the simpler case of the collapse and spin-up of an initially  
non-rotating core due to Hall effect.

\section{Weak Magnetic Fields and Potential Disk Formation}
\label{weakfield}

We now consider cases with a weak initial magnetic field of 
$B_0=10.6\muG$ (at an initial molecular hydrogen density of 
$n_{H_2}=10^5\cm^{-3}$), which is $30\%$ of the reference value. It
corresponds to a dimensionless mass-to-flux ratio of $\lambda = 9.73$ 
for the core as a whole, and $\lambda_{\rm c}=14.6$ on the central 
magnetic flux tube. If the dense core were to condense more or less
isotropically out of a more diffuse material of $10^{3.5}\cm^{-3}$ 
in density under the flux freezing condition, it would require only an  
unrealistically weak field of $1.06\muG$ in the diffuse gas for the 
core to have $B_0=10.6\muG$. For this reason, we believe that this 
value of $B_0$ is probably as low as, if not lower than, the minimum 
field strength that can be reasonably expected in the type of 
dense cores under consideration. 

Weaker fields are expected to be less efficient in magnetic braking. 
This is because the braking rate involves the product of the toroidal
and poloidal field strengths, and thus generally scales with 
the field strength as $B^2$. Indeed, for Model WREF which has the
same parameters as the reference model (Model REF) except for a 
weaker field of $B_0=10.6\muG$, a small ($\sim 20\AU$) 
rotationally supported disk is formed temporarily early in the
protostellar accretion phase, around the time $t\sim 3.68\times 10^{12}\second$, 
when the central mass is only $\sim 0.07\Msun$. This disk 
disappears at later times, however, because of 
strong magnetic braking, which drives a powerful outflow before the 
disk disappears (see Shang et al., in preparation).  

\begin{figure}
\epsscale{1.2}
\plottwo{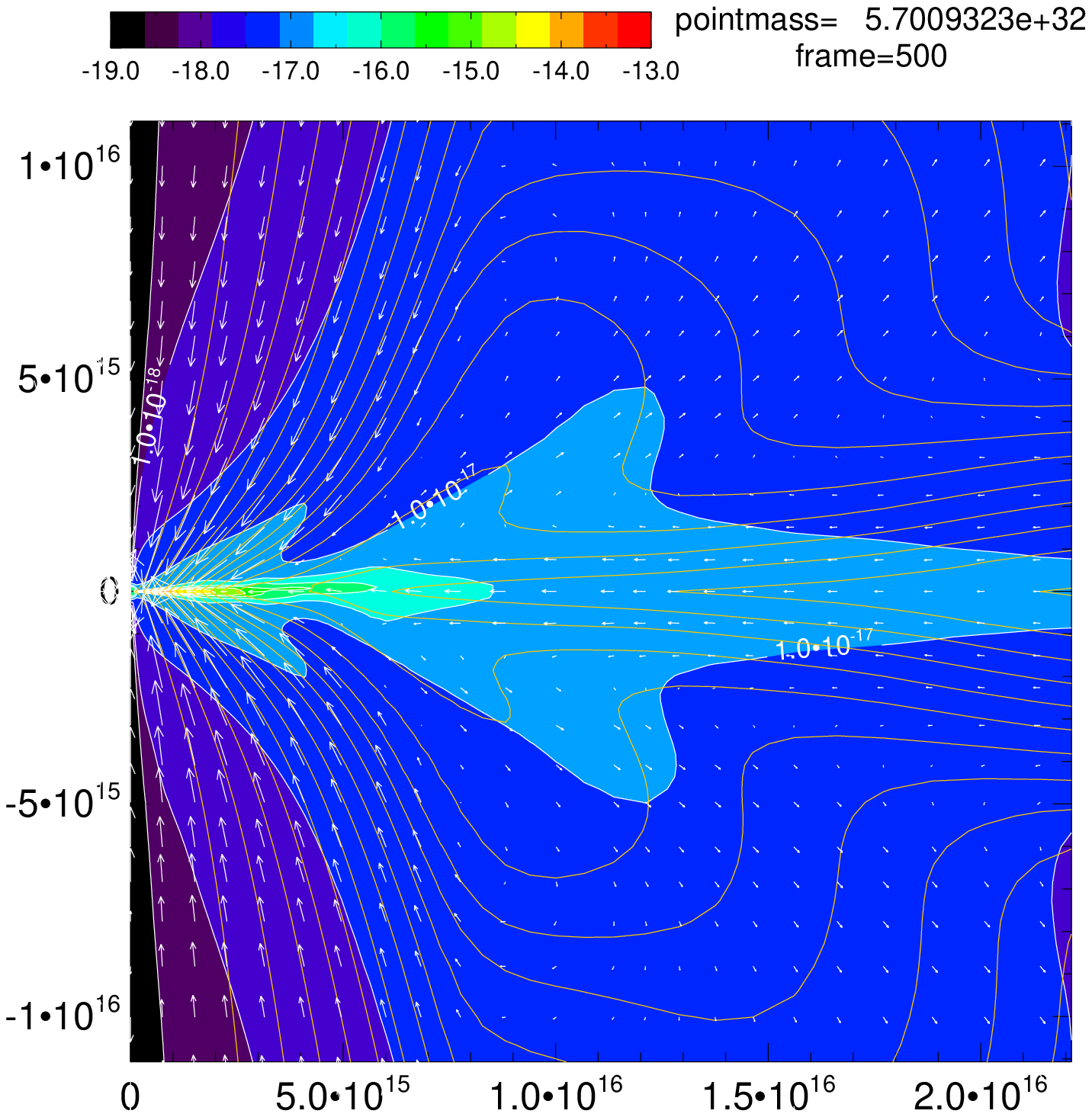}{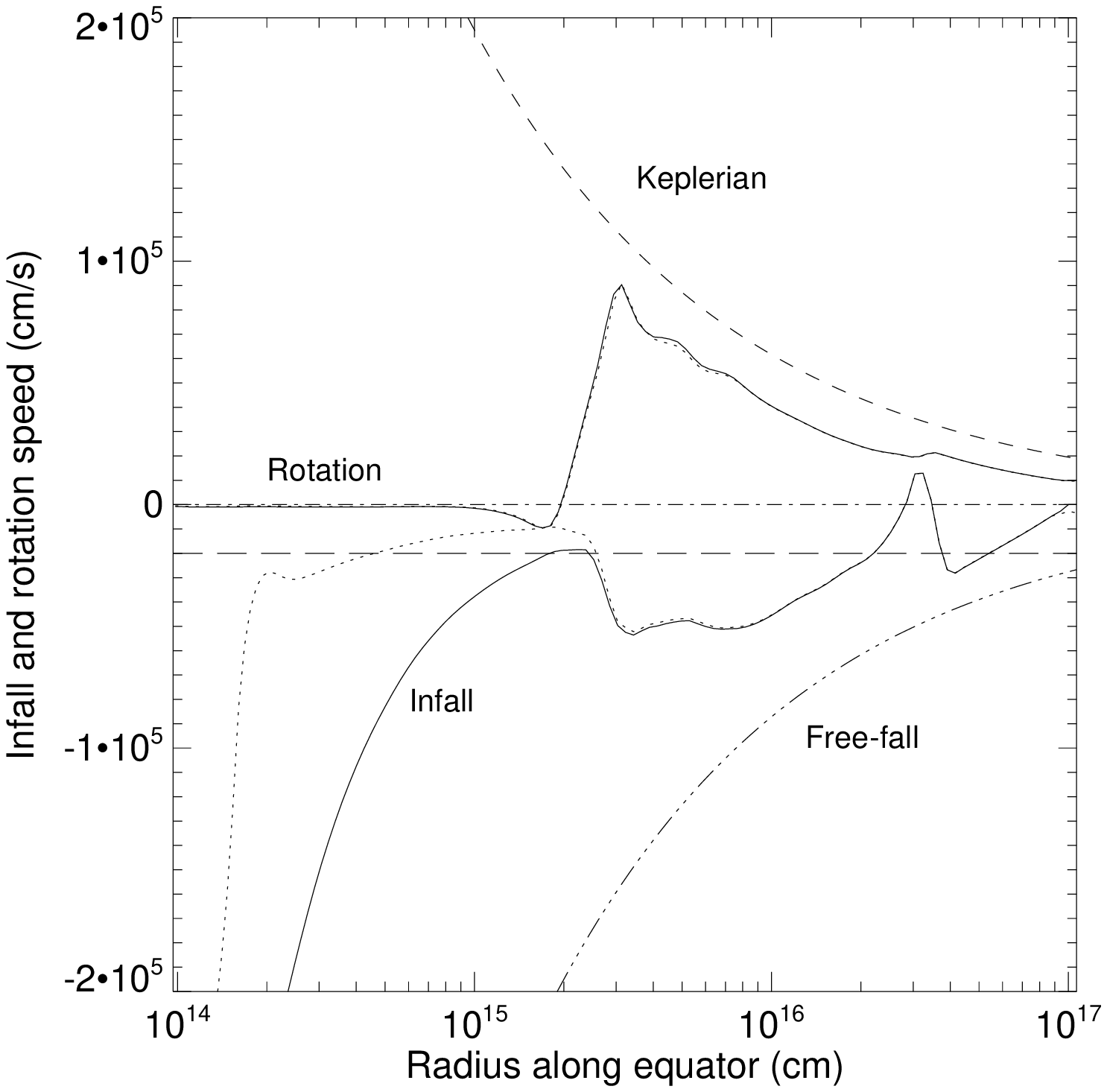}
\caption{Density color map, velocity field (white arrows) and magnetic 
field lines (yellow) of the collapsing core (left panel) and equatorial
 infall and rotation speeds (right panel) for the standard weak 
magnetic field model (Model WREF) at a relatively late time 
$t=5\times 10^{12}\second$, showing that the RSD formed earlier has by now
completely disappeared. Also plotted for 
comparison in the right panel are the Keplerian 
speed (upper dashed) and free fall speed (lower dash-dotted) 
based on the central mass, the sound speed (horizontal dashed 
line), and zero speed line. The ``effective ion speeds'' 
defined in eq.\ (4) are shown as dotted lines.  
 } 
\label{WeakB_NoDisk}
\end{figure}

The suppression of the rotationally supported disk at later times is 
illustrated in Fig.\ \ref{WeakB_NoDisk}. It includes a snapshot of 
the collapsing core (left panel) and a plot of the equatorial infall 
and rotation speeds (right panel), at $t=5\times 10^{12}\second$ when 
the central mass is $0.29\Msun$.  At this time, the prominent 
polar outflow at the 
earlier times has disappeared. It is replaced by a polar region of strong
infall, mostly along the magnetic field lines. The lack of outflow 
indicates the rotating disk that drives the outflow no longer exists. 
The disk suppression is shown clearly in the equatorial rotation
speed, which is close to zero inside the AD shock (around 
$10^{15}\cm$). The rapid, supersonic infall at small radii further 
supports the lack of an RSD, as in the more strongly magnetized 
standard model (see Fig.\ \ref{ReferenceModel} and \ref{pressures}). 
  
\begin{figure}
\epsscale{0.45}
\plotone{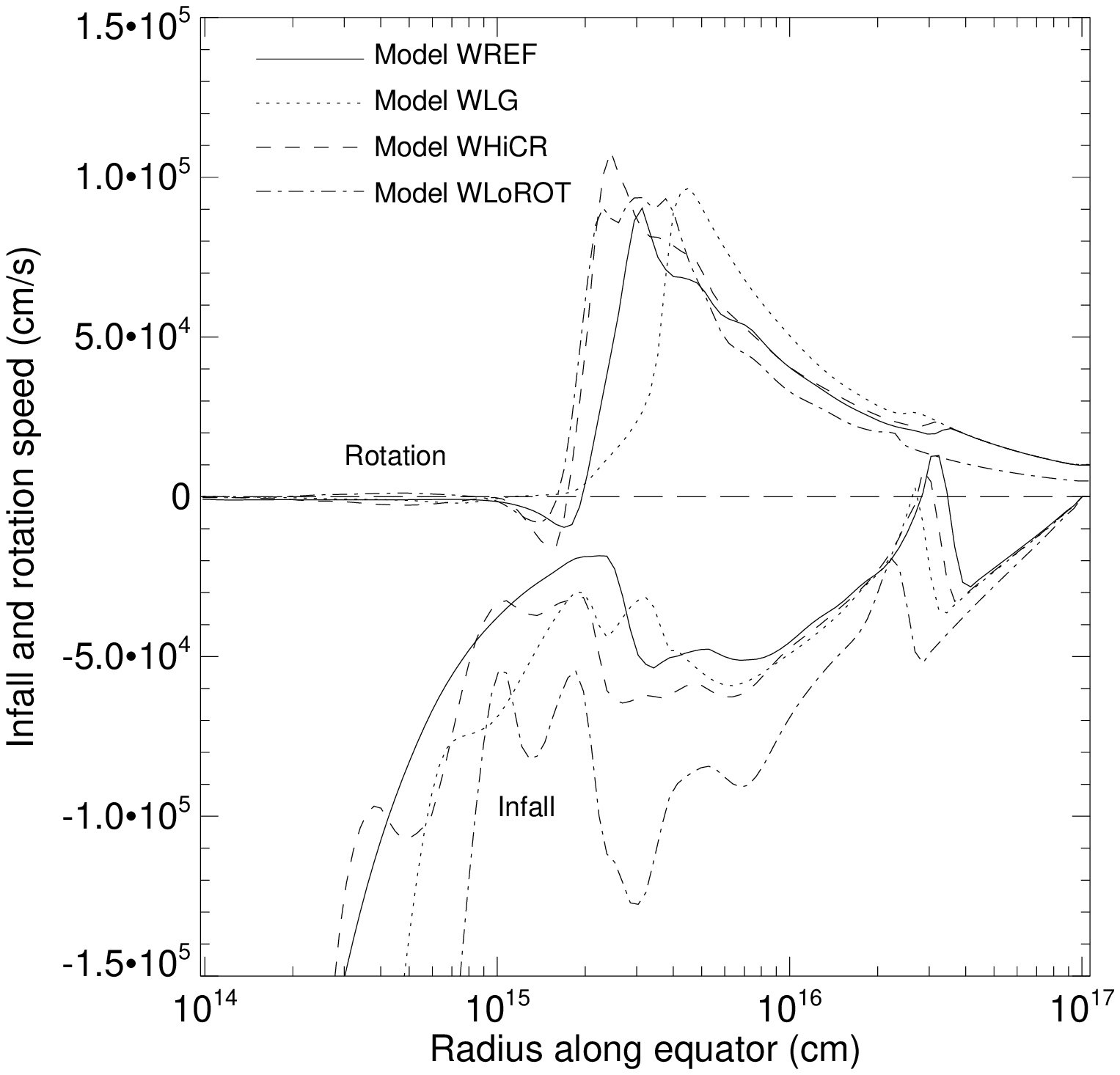}
\plotone{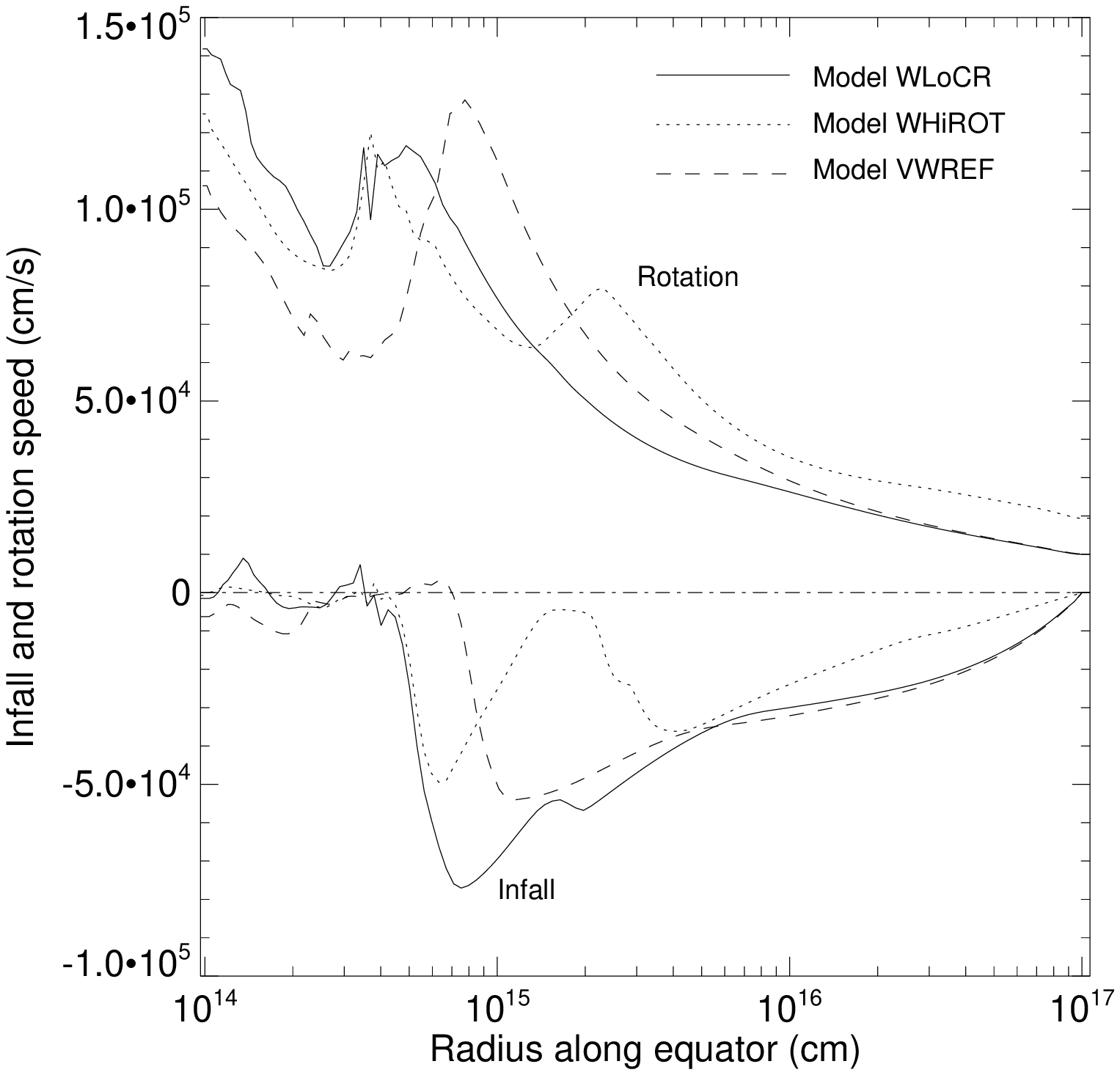}
\caption{Left panel: Infall and rotation speeds on the equator for 
those models with a weak magnetic field that do not have a 
rotationally supported disk at a relative late time $t=5\times 
10^{12}\second$. Right panel: Same as the left panel but for 
those models that do have what appear to be the rotationally supported 
disks at early times, when the central mass is only $\sim 1\%$ of 
the core mass. Whether these (small) disks can survive to much 
later times or not is unclear.   
}
\label{WeakB_others}
\end{figure}

Disk suppression is not unique to Model WREF. In the left panel of 
Fig.\ \ref{WeakB_others} we show three additional examples (Models
WLG, WHiCR and WLoROT in Table 1), where the disk formation is 
suppressed at a relatively late time $t=5\times 10^{12}\second$, when 
the central mass is 
$0.40$, $0.34$ and $0.65\Msun$, respectively. These models have 
the same weak magnetic field as Model WREF, but have either a larger 
grain size (Model WLG), a higher cosmic ray ionization rate (Model
WHiCR), or a lower initial rotation rate (Model WLoROT). The lack of 
disk is not too surprising for 
Models WLG and WHiCR, because the magnetic fields are better
coupled to the neutral matter in these models than in Model 
WREF. The absence of a disk in Model WLoROT is also expected, 
because its more slowly rotating core is more easily braked. 

Some rotationally supported disks do form with other choices of 
parameters, however, at least at early times. The disk 
formation is illustrated in 
the right panel of Fig.\ \ref{WeakB_others}, where the equatorial 
infall and rotation speeds for Models WLoCR (same as the standard 
weak field model but with a lower cosmic ray ionization rate), 
WHiROT (with a higher initial core rotation rate) and VWREF (with 
an unrealistically low $B_0=3.54\muG$) 
are plotted at a early time, when the central mass is only $1.79\times
10^{-2}$, $1.27\times 10^{-2}$ and $1.05\times 10^{-2}\Msun$,
respectively. There is more mass in the rotationally supported 
disk than at the center, which is why the rotation curve is 
non-Keplerian except close to the origin. Each of the disks drives a 
strong, sometimes chaotic, outflow, which makes it hard to continue 
the non-ideal MHD simulation reliably to much later times. The 
disk in the weakest 
field case (Model VWREF) evolves into a ring, which may fragment in 
3D. It is unclear whether the early disk in the other two cases can 
survive subsequent magnetic braking or not (it did not in the 
standard weak field model WREF). Once a disk becomes 
self-gravitating, gravitational torque will likely become important 
in the disk dynamics. This important effect is not captured in 
our axisymmetric simulations. 
In any case, it is clear that disk formation in the moderately weak 
magnetic field case is more complicated than the moderately strong 
field case, with the outcome depending on the core rotation rate 
and the degree of field-matter coupling, at least at early times. 
Paradoxically, the weaker field cases are more difficult to 
simulate because of strong, chaotic outflows. More work is needed 
before firmer conclusions on disk formation can be drawn.

\section{Hall Spin-up of Non-Rotating Envelope}
\label{PureHall}

Of the three non-ideal MHD effects, the Hall effect is the least
explored in the context of core collapse and disk formation (see, 
however, \ct{Krasnopolsky_2011} and \ct{Braiding2011}). We have
seen in \S{}\ref{Hall}  that it can spin up the nearly completely 
braked post-AD shock material in an initially rotating core to a 
significant speed, and the sense of the Hall-induced rotation 
depends on the orientation of the magnetic field. This Hall 
spin-up can be illustrated even more clearly in the collapse of 
an initially non-rotating core, where any rotation that develops 
subsequently must come solely from the Hall effect. 

\begin{figure}
\epsscale{0.45}
\plotone{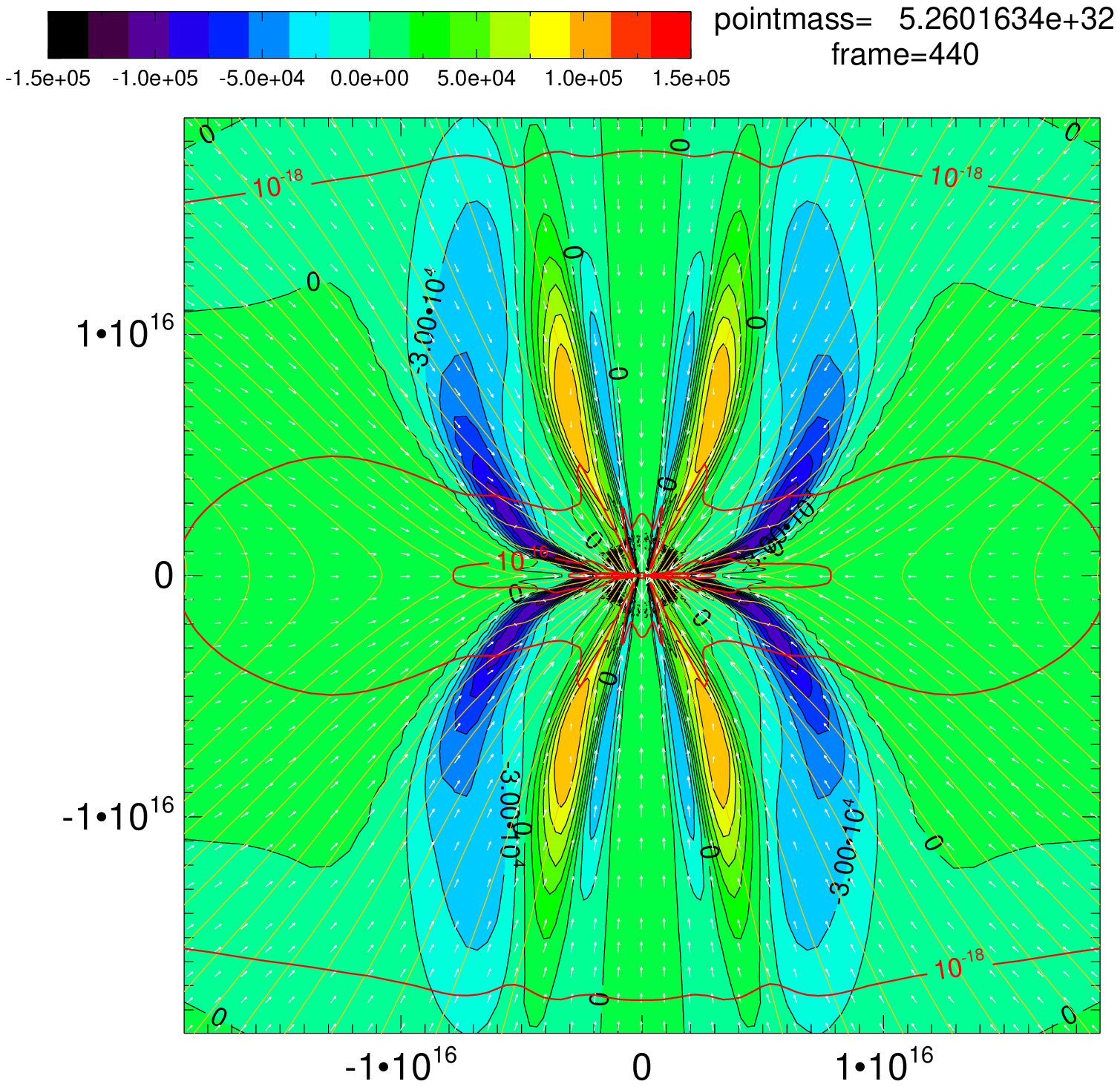}
\plotone{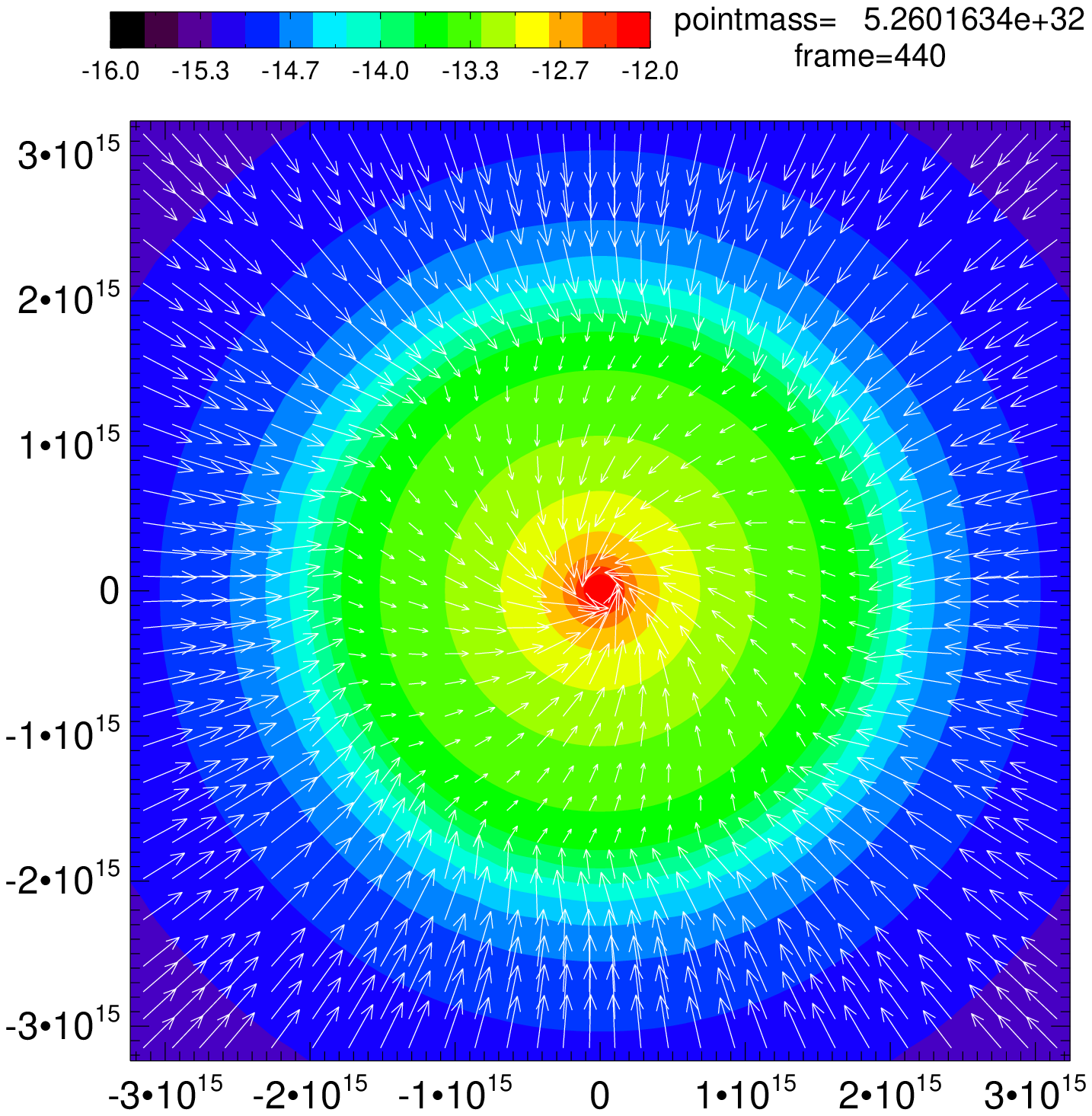}
\caption{Left panel: map of the Hall-induced rotation speed during 
the collapse of an initially non-rotating core (Model REF$_{AHO}^-$) at a 
representative time $t=4.4\times 
10^{12}\second$, when the central mass is $0.26\Msun$. Also plotted  
are magnetic field lines (yellow), isodensity contours (red, spaced by
a factor of 10), and velocity vectors (white). Right panel: map of 
$log(\rho)$ (color) and velocity field (white vectors) in the equatorial 
plane, showing the spin up of the dense, post-AD shock region due to
Hall effect. 
}
\label{PureHall_vphi}
\end{figure}

We will concentrate on Model NoROT$_{AHO}^-$ where all three non-ideal
MHD effects are included and the initial magnetic field
is anti-parallel to the rotation axis. We have confirmed
that the case with opposite field orientation (Model NoROT$_{AHO}^+$) 
produces identical results, except
that the sign of the Hall-induced rotation is flipped, as expected. 
The left panel of Fig.\ \ref{PureHall_vphi} gives an overall impression of the
Hall-induced rotation on an intermediate scale of $\sim 10^{16}\cm$,
at a representative time $t=4.4\times 10^{12}\second$, when the central 
mass is $0.26\Msun$. Note the alternating pattern of negative
and positive rotation speeds, with maximum values reaching $\sim 
10^5\cm\second^{-1}$, which is much higher than the sound speed ($2\times 
10^4\cm\second^{-1}$). On smaller scales, the structure is dominated 
by a dense flattened pseudodisk that is visible from the iso-density
contours in Fig.\ \ref{PureHall_vphi}. 

\begin{figure}
\epsscale{0.45}
\plotone{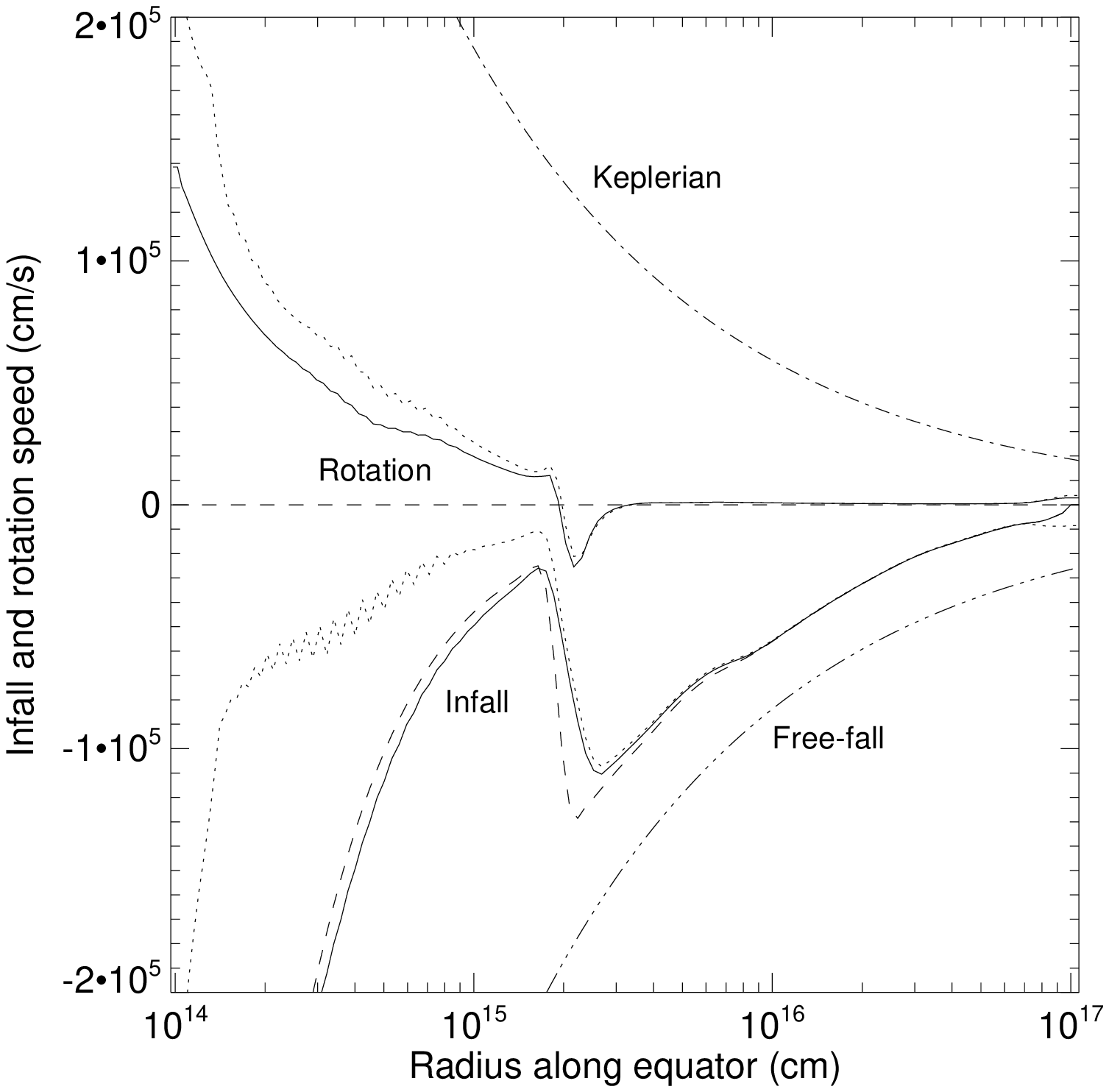}
\plotone{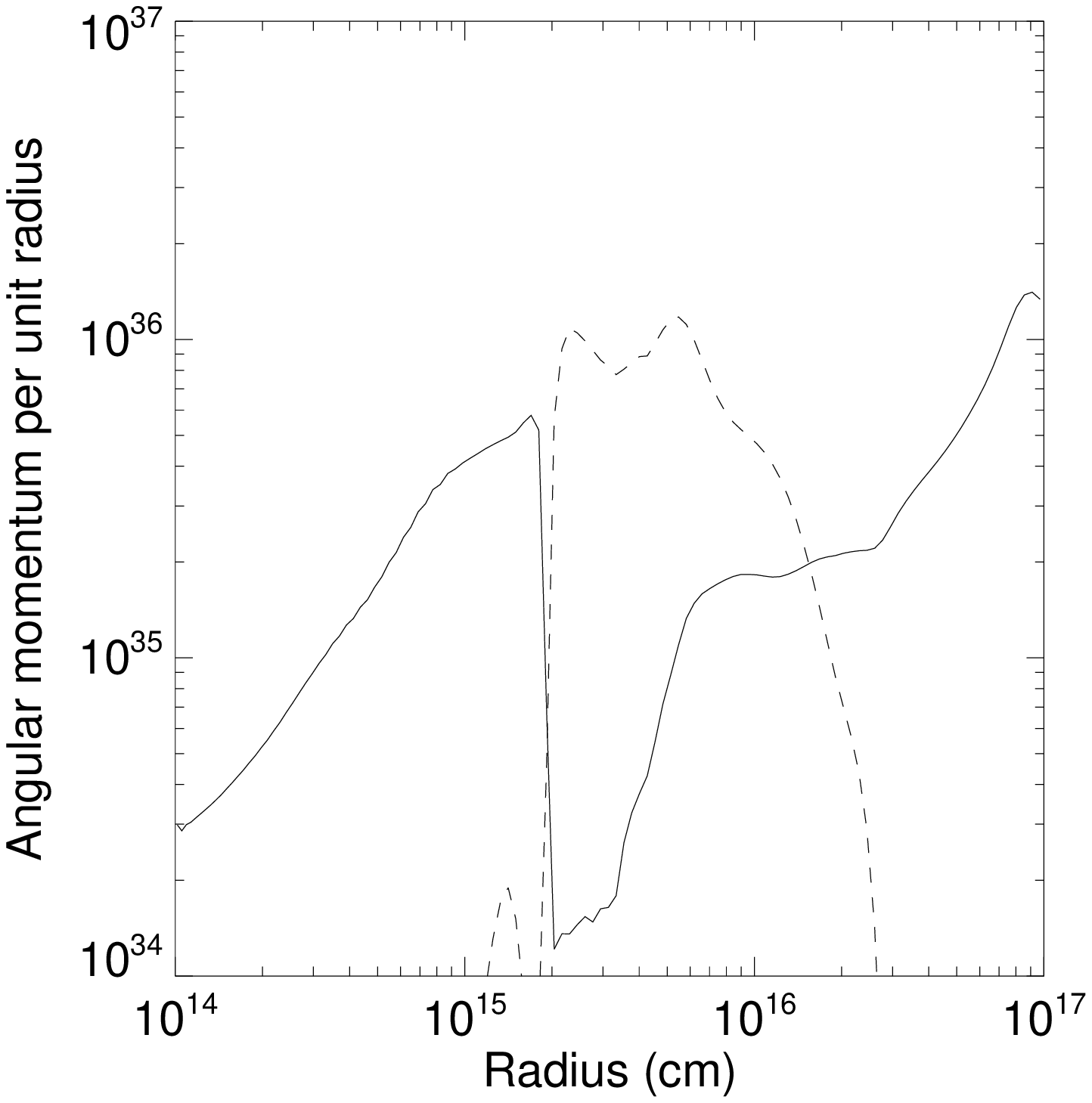}
\caption{Left Panel: equatorial infall and rotation speeds of 
the neutral matter (solid lines) for the collapse of an initially
  non-rotating core. Also plotted for 
comparison are ``effective ion speeds'' (defined in equation [4],
dotted lines), and the infall and rotation speeds in an AD-only 
case that does not have any Hall 
effect (dashed lines). Right panel: Positive (solid line) and negative
(dashed) angular momentum per unit radius $dL/dr$ as a function of 
radius.  
}
\label{PureHall2}
\end{figure}

The dense equatorial pseudodisk is collapsing as well as spinning in 
the positive azimuthal direction, as shown pictorially in the right 
panel of Fig.\ \ref{PureHall_vphi} and more quantitatively in the left 
panel of Fig.\ \ref{PureHall2}. The collapse shows rapid deceleration 
near $\sim 2\times 10^{15}\cm$ and reacceleration interior to it, 
characteristic of an AD shock. The infall speed is similar to that in 
the pure-AD case, indicating that the Hall effect modifies relatively 
little the overall collapse dynamics. The difference in rotation speed
is more pronounced, especially in the post-shock region, where the 
maximum rotation speed exceeds $10^5\cm\second^{-1}$, comparable to the peak value 
on the large scale shown in Fig.\ \ref{PureHall_vphi}. The rotational 
component of the ``effective 
ion speed'' (defined in equation [4]) is larger than the neutral
rotation speed, indicating that a magnetic force is exerted in the 
positive azimuthal direction to drive the ion-neutral drift. It is 
the same force that torques up the pseudodisk. The force is 
particularly large in the postshock region, because the magnetic 
field is strong there. Nevertheless, the spin up fails far short 
of reaching a Keplerian rate, which is the reason why the overall 
collapse dynamics is little affected.  

In an initially non-rotating core, any spin up in one direction must
be offset by a spin up in the opposite direction, so that the total
angular momentum is conserved. In the region that extends from the
origin up to the AD shock, the net angular momentum is positive (see 
the right panel of Fig.\ \ref{PureHall2}), because it is dominated 
by the positively spinning 
pseudodisk. Outside the AD shock, the net angular momentum is
negative, dominated by the hour-glass shaped counter-rotating region
shown in left panel of Fig.\ \ref{PureHall_vphi}. One may expect 
the positive and 
negative angular momenta to sum up to zero over the entire
computational volume. However, this is not the case, because the total
angular momentum is dominated by the material at large distances 
(near the core edge) whose rotation speed is small but non-zero 
(see the left panel of Fig.\ \ref{PureHall2}). Because the mass 
at larger distances is larger and 
has a longer lever arm, it dominates the total angular momentum. 
The total positive angular momentum inside the computation domain 
is $6.52\times 10^{52}\gram\cm^2\second^{-1}$ whereas the total 
negative angular momentum is $-9.38\times 10^{51}\gram\cm^2\second^{-1}$. 
They do not cancel out exactly. Some of the angular momentum must have
left the computation box, through torsional Alfv{\'e}n waves. Dividing the
total net angular momentum by the total mass left in the simulation
box ($1.47\times 10^{33}\gram$) yields $3.80\times
10^{19}\cm^2\second^{-1}$, which is the average specific angular
momentum. For a core size of $10^{17}\cm$, the corresponding
characteristic rotation speed is $3.8\times 10^2\cm\second^{-1}$, 
about $2\%$ of the sound speed, much smaller than the rotation speed
achieved on the $10^{16}\cm$ scale or smaller. We conclude that, 
despite the localized, supersonic rotation induced on small scales, 
the influence of the Hall effect on the global dynamics is limited.    

\section{Discussion}
\label{discussion}

\subsection{Why is Protostellar Disk Formation Difficult in 
Magnetized Cores?}

\subsubsection{Ideal MHD Limit}
\label{imhd}

The fundamental reason for the difficulty in forming protostellar 
disks in magnetized cores is that the protostellar collapse 
concentrates magnetic flux at small radii, precisely where the 
rotationally supported disk (RSD) tends to form in the absence of 
magnetic braking. The basic difficulty can be seen most clearly 
in the ideal MHD limit, where magnetic flux is dragged into the
central star (because of flux freezing) to form a split magnetic 
monopole. The rapid increase in field strength towards the 
central object enables the (split) monopolar field to brake 
the circumstellar disk catastrophically, as shown analytically 
by \citeauthor{Galli_2006} (\citeyear{Galli_2006};
see their Fig.\ 1 for a sketch of the 
expected field geometry). 

A case can also be made for catastrophic disk braking in the ideal MHD limit 
from numerical simulations (\ct{Allen_2003}; \ct{MellonLi2008};
\ct{HennebelleFromang2008}), although it is less clean cut. The reason is 
that, as the mass of the protostar grows, more and more magnetic 
flux is dragged to the origin, which creates a stronger and 
stronger (split) magnetic monopole that squeezes more and more 
strongly on the material on the equator from above and below. 
When the oppositely directed magnetic fields above and below 
the equator are squeezed within a single cell of each other, 
numerical reconnection becomes unavoidable. The expected 
numerical reconnection is present in the ideal MHD simulations 
of \citet{MellonLi2008}, especially for relatively strongly 
magnetized cores (with a dimensionless mass-to-flux $\lambda$ 
of several or less; see their Figs.\ 14-16), where no RSDs 
form but the numerical results are complicated by reconnections. 
The collapse of more weakly magnetized cores 
of $\lambda \sim 10$ is not significantly affected by 
reconnections, and yet no RSDs form either. The conclusion 
from the ideal MHD simulations is that magnetic braking is 
efficient enough to suppress disk formation for $\lambda \lesssim 
10$. 

The above conclusion is strengthened by the resistive MHD 
disk-formation calculations of \citet{Krasnopolsky_2010}, 
which included a wide range of (prescribed) resistivity. They 
found that, as the resistivity is decreased below a certain 
value, (numerical) reconnection starts to become important 
(as expected), which complicates the interpretation of the 
numerical results. However, before the reconnection sets in, 
the RSD is already completely braked by a moderately strong 
magnetic field. Extrapolating the results for those clean, 
low resistivity runs without numerical reconnection to the zero 
resistivity limit (where numerical reconnection is 
unavoidable) indicates that complete disk suppression also 
holds true for the ideal MHD case. 

\citet{Machida_2010} appears to have come to a different conclusion. 
They found a $10^2$-AU scale disk in their 3D nested-grid 
simulations in the ideal MHD limit (their Model 4), even 
though their core is strongly magnetized (with a global 
dimensionless mass-to-flux ratio of $\lambda=1$, see their 
Table 1). It is unclear whether the disk is rotationally 
supported or not. If yes, the result would be hard to
understand. For such a strongly magnetized core, the mass 
accumulation at the protostar should produce a strong 
(split) magnetic monopole in the ideal MHD limit, which 
is expected to trigger powerful numerical reconnection, 
as  discussed above and shown in \citet{MellonLi2008}; we 
have re-run their Model 4 and found the expected episodic 
reconnections and no rotationally supported disk. There 
were no reconnections mentioned in their paper, and the 
apparent lack of reconnection may be an indication that 
considerable numerical diffusivity acts to reconnect magnetic 
field lines of opposite polarities efficiently and prevent 
magnetic flux from accumulating near the protostar to form 
the expected split monopole in the first place. Since the 
trapping of magnetic flux at small radii lies at the heart 
of the efficient braking that renders disk formation difficult 
in the ideal MHD limit, inability to do so numerically may 
weaken the braking efficiency artificially and lead to disk 
formation.  
   
\subsubsection{Non-Ideal MHD Effects} 

In lightly ionized dense cores of molecular clouds, non-ideal 
MHD effects are to be expected. Non-ideal effects, particularly 
ambipolar diffusion and Ohmic dissipation, enable the bulk 
neutral matter to move across magnetic field lines, breaking 
the flux freezing condition that is responsible for the 
formation of the central split magnetic monopole which, in 
turn, is responsible for the catastrophic disk braking in 
the ideal MHD limit. The elimination of the central split 
monopole does not necessarily mean, however, that the magnetic 
braking would automatically be weakened enough for an RSD 
to appear. The reason is that the magnetic flux that would 
be trapped in the central split monopole in the ideal MHD 
limit is now concentrated in a small, but finite, 
circumstellar region instead, as first demonstrated by 
\citet{LiMcKee1996} in the case of ambipolar diffusion (AD). 
\citet{KrasnopolskyKonigl2002} showed semi-analytically 
that the AD-induced flux concentration at small radii can 
in principle suppress disk formation completely, just as 
in the ideal MHD limit. \citet{MellonLi2009} showed 
numerically that RSDs are indeed suppressed by a moderately 
strong magnetic field (with $\lambda \sim$ several) in 
the presence of ambipolar diffusion for a reasonable range
of cosmic ray ionization rate. 

The result of \citet{MellonLi2009} is strengthened by the 
simulations presented in \S{}\ref{PureAD}. We improved 
over their calculations by self-consistently computing 
charge densities including dust grains and by extending 
the computation to the prestellar phase of core evolution 
leading up to the central mass formation. The extension 
allows us to explore the angular momentum evolution and 
disk formation during the transition between the 
prestellar and protostellar phase of star formation.   
We find that the tendency to form a disk is stronger 
around the time of initial protostar formation than at 
later times (see Fig.\ \ref{transition}). This is because 
there is as yet little magnetic flux accumulated near 
the center and it takes time for the magnetic braking 
to remove angular momentum. Once enough magnetic flux
has accumulated near the protostar to drive a 
well-developed accretion (C-)shock, the field strength 
in the post-shock region is typically strong enough to 
remove most of the angular momentum of the material 
falling into the central object, as long as the dense 
core is moderately strongly magnetized to begin with 
(with $\lambda \sim$ a few to several). 

The conclusion that the RSD is suppressed by a moderately 
strong magnetic field in the presence of AD is robust, 
because the size of the AD shock and the post-shock field 
strength are rather insensitive to cosmic ray ionization 
rate and the grain size distribution (see Fig.\ \ref{general}). 
They are determined mostly by the global requirements that 
(1) most of the magnetic flux associated with the central 
stellar mass be redistributed in the post-shock region, and 
(2) the strong postshock magnetic field be confined by the 
infall ram pressure (see equations [8] and [10] of \ct{LiMcKee1996}),
as long as the width of the C-shock is small 
compared to the radius of the shock. In principle, if the 
ionization level is decreased by a arbitrarily large factor, 
the magnetic field would eventually decouple completely from 
the bulk neutral material and an RSD would form. In practice, 
however, the RSD is suppressed even for the highly conservative 
case of both an unrealistically low cosmic ray ionization 
rate of $\zeta = 10^{-18}\second^{-1}$ and an MRN grain size 
distribution that contains a large amount of small grains 
(Model LoCR in Table 1). Both the grain growth expected in 
dense cores and a more realistic (higher) cosmic ray 
ionization rate tend to make the magnetic field better 
coupled to the bulk neutral material and the magnetic 
braking more efficient.
  
The Ohmic dissipation does not change the above picture much, because 
the Ohmic diffusivity $\eta_O$ is well below the ambipolar diffusivity 
$\eta_{AD}$ for the density range $n_H \lesssim 10^{12}\cm^{-3}$
that is crucial for disk formation. We can estimate the ratio of the 
two diffusivities through 
\begin{equation}
\label{etao_etaad}
{\eta_O\over \eta_{AD} } \sim {n_c \over n_e} {1\over \beta_c \beta_e}
\end{equation}
where the subscripts ``$c$'' and ``$e$'' denote, respectively, the 
charged species whose contribution dominates the AD term and the 
electrons that are mainly responsible for the Ohmic term. 
The (dimensionless) Hall parameter $\beta\equiv \tau \omega$ 
(where $\tau$ is the collisional damping time of the motion 
of a charged species relative to the neutral and $\omega$ the 
cyclotron frequency) provides a measure of how well a charge 
is tied to the magnetic field. The relative unimportance of 
Ohmic dissipation comes mainly from the fact that electrons 
are extremely well tied to the magnetic field in the density 
regime of interest, with $\beta_e \gg 1$. For example, at a 
representative density of $n_H=3\times 10^8\cm^{-3}$, we 
find $\beta_e \approx 2\times 10^5$ for the field 
strength-density relation given 
by equation (5). In the large grain (LG) case that we have  
considered, the AD term is dominated by metal ions, $M^+$. 
Their number density is close to the electron number density 
($n_{M^+} \approx n_e$) and they are well tied to the magnetic
field at the representative density, with $\beta_{M^+}\approx 
40$. From equation (\ref{etao_etaad}), we expect $\eta_O/\eta_{AD} \sim 
10^{-7}$, which is in agreement with the computed values 
shown in the right panel of Fig.\ \ref{Diffusivities}. In the 
case of MRN grain size distribution, the AD term is dominated 
by the small negatively charged grains, $g^-$. At the 
representative density, there are about $10^2$ small negatively 
charged grains for each electron ($n_{g^-} \approx 10^2 n_e$) 
and the small charged grains are marginally tied to the 
magnetic field $\beta_{g^-}\sim 1$. Both the smaller electron 
abundance and the weaker coupling of the grains to the field 
tend to make Ohmic dissipation more important relative to 
ambipolar diffusion. Nevertheless, the electrons are so well
tied to the magnetic field that, even in this case, the Ohmic 
diffusivity is still much smaller than the ambipolar diffusivity, 
by a factor of $\sim 2000$ at the representative density. This 
estimate is again in agreement with the computed values shown 
in the left panel of Fig.\ \ref{Diffusivities}.  

The Hall effect is expected to be more important than Ohmic 
dissipation in diffusing the magnetic field in the density 
regime under consideration. This can be seen from the ratio 
of Hall and ambipolar diffusivities 
\begin{equation}
{\eta_H\over \eta_{AD} } \sim {1\over \beta_c},
\end{equation}
which is applicable under the conditions that the same charged species 
``$c$'' dominates both the Hall and AD terms and $\beta_c \gtrsim 
1$. These conditions are satisfied for the large grain (LG) case  
at the representative density $n_H=3\times 10^8\cm^{-3}$, where 
metal ions dominate both terms and $\beta_{M^+}\approx 40$. In this 
case, the Hall diffusivity is only about $2\%$ of the ambipolar 
diffusivity (but still much larger than the Ohmic diffusivity). It 
becomes comparable to the ambipolar diffusivity in the case of MRN
grain size distribution, where both terms are dominated by small 
dust grains that are marginally coupled to the magnetic field 
(with a Hall parameter $\beta$ of order unity) at the representative
density. At higher densities, the Hall diffusivity is expected 
to exceed the ambipolar diffusivity as small grains become 
even less well tied to the magnetic field. However, the 
contribution from positively charged grains start to cancel out 
that from negatively charged grains, leaving the Hall diffusivity
comparable to the ambipolar diffusivity over a wide range of 
density (see the left panel of Fig.\ \ref{Diffusivities}). The Hall 
effect therefore does not increase the magnetic diffusivity by more
than a factor of a few. As such, it is not expected to 
greatly change the global flow dynamics, especially the structure 
of the ambipolar diffusion-induced accretion shock, which 
lies at the heart of the magnetic braking catastrophe. It does, 
however, introduce a new ingredient into the problem: it can 
{\it actively} torque up a magnetized collapsing envelope, even 
if the envelope is non-rotating to begin with, as first 
pointed out by \citet{WardleNg1999} and demonstrated numerically 
in \citeauthor{Krasnopolsky_2011} (\citeyear{Krasnopolsky_2011};
see also \ct{Braiding2011}). On 
the scale of several AUs or larger that we can resolve in our 
non-ideal MHD simulations (\S{}\ref{Hall} and \ref{PureHall}), the 
Hall spin-up does not reach the Keplerian speed. 
The angular momentum gained through Hall spin-up may, however, be 
conducive to the formation of RSDs on smaller scales, particularly 
at high enough densities where electrons begin to decouple from the 
magnetic field and Ohmic dissipation becomes the dominant 
process for field diffusion (\ct{Machida_2010}; \ct{DappBasu2010}).
Nevertheless, the problem of catastrophic magnetic braking 
that prevents the formation of a sizable RSD of tens of AUs or 
larger is not resolved through the three non-ideal MHD effects.  

\subsection{Limitations and Future Directions: How to Form RSDs?}

We are unable to produce robust, large-scale, rotationally supported 
disks in our non-ideal MHD simulations for dense cores magnetized 
to a realistic level. And yet, rotationally supported disks are 
observed around young stars, at least at relatively late times, 
after the massive envelope has been removed (which reveals the 
embedded disk for direct rotation measurement; see \ct{WilliamsCieza2011}
for a review). Clearly, one or more assumptions made 
in our calculations must break down in order for the observed 
(late-time) RSDs to form. These include (1) assumptions made in 
the setup of the numerical problem because of computational 
constraints and (2) additional physical effects that we have not 
taken into account. We comment on these limitations and their 
relevance to RSD formation in turn. 

\subsubsection{Numerical Limitations and Possible Ways to Form RSDs}
\label{NumericalLimitations}

We have restricted our problem setup to 2D (assuming axisymmetry), 
which greatly reduced the computational demand. A potential 
drawback is that the imposed symmetry may have enhanced the 
ability of the AD shock in trapping magnetic flux near the 
central object, which lies at the heart of the catastrophic 
magnetic braking that prevents disk formation. The reason is 
that the high magnetic pressure may force the trapped field 
lines to escape along the path of least resistance when the 
axisymmetry is broken. It is plausible that some magnetic flux 
loaded with relatively little matter (recall that the post-shock
region is magnetically subcritical and supported against free-fall 
collapse by magnetic forces) would act as 
a ``light'' fluid and escape in some azimuthal directions, 
allowing less magnetized fingers of ``heavy'' material to 
sink closer to the center in other directions. This type of 
interchange instability was considered in \citet{LiMcKee1996}.
It may weaken the magnetic field strength at small 
radii enough to enable disk formation. Investigation of this 
possibility is now underway. 
 
Another limitation of the current problem setup is that we adopted a 
relatively large central hole around the protostar (typically 
$10^{14}\cm$ or $6.7\AU$ in radius). We have experimented with smaller 
holes by a factor of 2--3 in a few cases and found quantitatively
similar results. However, it is difficult to reduce the hole 
size by a much larger factor, because of the constraints on the 
time step, which decreases as the square of the hole size (assuming
the same resolution in the $\theta$-direction) in our explicit 
treatment of the non-ideal MHD terms. A drawback is that we are 
unable to determine whether small, (sub-)AU-scale RSDs can 
form during the main protostellar accretion phase or not. The 
existence of such small RSDs is suggested by the powerful molecular 
outflows ubiquitously observed around deeply embedded, Class 0 
protostars (e.g., \ct{Bontemps_1996}); they are generally 
thought to be driven by a fast primary wind launched 
magnetocentrifugally from the inner part of a Keplerian disk 
close to the central object (\ct{Shu_2000}; \ct{KoniglPudritz2000}). 

Small RSDs can in principle form at high enough densities where 
electrons start to decouple from the magnetic field. Before 
thermal ionization of alkali metals becomes important, the 
Ohmic dissipation can reduce the local current density, making 
it hard for the magnetic field to bend in the poloidal plane 
(which limits the flux accumulation at small radii) and to 
twist in the azimuthal direction (which weakens magnetic 
braking; \ct{Shu_2006}; \ct{Machida_2007}; \ct{Krasnopolsky_2010};
\ct{DappBasu2010}). A worry is that, in the 
presence of only ambipolar diffusion and classical Ohmic dissipation,
our (2D axisymmetric) 
calculations showed that little angular momentum is left at 
the inner edge of the computation domain ($6.7\AU$), making the 
formation of RSDs interior to it difficult. However, the Hall
effect can spin up the flow that collapses through the inner
boundary to a supersonic (although still locally sub-Keplerian) rotation 
speed, perhaps making the formation of small RSDs possible. 
Alternatively, the 3D magnetic interchange instability discussed 
earlier may weaken the magnetic braking interior to the AD
shock enough to facilitate the formation of RSDs in general, 
and the small RSDs required for fast wind launching in 
particular. When and how the small RSDs grow to large RSDs 
observed around relatively evolved YSOs remains uncertain, 
and may require additional physical effects that have not 
been investigated in detail in this context to date.

\subsubsection{Additional Physical Effects for RSD Formation}

An important ingredient for low-mass star formation is protostellar 
outflow. It may play a crucial role in disk formation. As first 
proposed by \citeauthor{MellonLi2008} (\citeyear{MellonLi2008};
see their \S{}6.2.2), the outflow 
can strip away the slowly 
rotating protostellar envelope, which brakes the equatorial infall
material that is magnetically linked to it and that tries to spin up 
and form a rotationally supported disk. Part of the envelope may be 
removed by the core collapse process itself (see, e.g., \ct{Machida_2010})
but, if the efficiency of star formation in individual 
low-mass cores is of order $1/3$ (\ct{Alves_2007}; \ct{Andre_2010}),
the majority of the envelope mass must be removed by 
some other process, most likely a (fast) protostellar wind \citep{MatznerMcKee2000}.
Indeed, the bipolar molecular outflow, thought to 
be primarily the envelope material swept up by a fast wind
(\ct{Shu_1991}; \ct{Shang_2006}), is observed to have a narrow 
jet-like appearance along the axis during the early Class 0 phase 
and the opening angle at the base increases as the YSO ages
\citep{ArceSargent2006}. 
As the fast wind sweeps out an increasingly wider polar region in the 
envelope, the braking efficiency of the remaining equatorial infall 
region should decrease, perhaps to a low enough value 
that a large-scale rotationally supported disk can form. 

A specific outflow-enabled large-scale disk formation scenario is as 
follow. We envision the early formation of a small (perhaps AU-scale) 
rotationally supported disk (unresolved by the current generation of
instruments) during the Class 0 phase, through the processes 
discussed in \S{}\ref{NumericalLimitations}. Although the small 
disk can grow gradually through internal angular momentum redistribution 
(perhaps gravitational torques rather than magnetic stresses 
since magnetic decoupling is required for the disk to 
form in the first place, see discussion in \ct{DappBasu2010}), we 
envision rapid growth in disk size (to, say, $100\AU$ or 
more) only during the late phase of envelope removal, when the 
braking of the equatorial infall material by the envelope is 
rendered inefficient by outflow stripping. This envelope-depletion 
induced rapid disk growth may occur towards the end of the main 
protostellar mass accretion phase, perhaps during the transition 
from the Class 0 to Class I phases of (low-mass) star formation. 
Detailed calculations and high resolution observations, perhaps 
using ALMA, are needed to test this scenario of late-time 
formation of a large-scale disk. 

Another possibility for large-scale RSD formation is through enhanced 
magnetic diffusivity. 
If the diffusivity is greatly enhanced over the classical microscopic 
values considered in this paper by some processes, the magnetic 
braking may be weakened enough to allow for RSD formation. 
\citet{Shu_2006} was the first to propose that enhanced Ohmic 
resistivity may enable RSD formation, and this was demonstrated 
explicitly in \citet{Krasnopolsky_2010}. The enhancement in 
resistivity may come from turbulence \citep{Kowal_2009},
which is observed in dense cores from nonthermal line width, or 
current-driven instabilities \citep{NormanHeyvaerts1985}, although 
these effects are hard to quantify at the present time (see, however, 
\ct{Santos-Lima_2010} who have started to quantify the  
so-called ``turbulent reconnection diffusivity''). Similarly, 
\citet{Krasnopolsky_2011} showed that if the Hall diffusivity is 
large enough, it can enable RSDs to form even in initially 
non-rotating dense cores. 

Numerical diffusion may mimic to some extent enhanced magnetic 
diffusion of physical origins and lead to large-scale RSD formation. 
\citet{Machida_2010} was able to produce large-scale disks in 
strongly magnetized cloud cores in both the ideal MHD limit and 
with a classical (microscopic) value of Ohmic dissipation. However, 
as we have argued in \S{}\ref{imhd}, the lack of episodic reconnections 
expected in the ideal MHD limit indicates a considerable numerical
diffusion in their calculations. Another indication is that, 
in the presence of only the classical Ohmic dissipation, we do 
not find any large-scale RSD (Model REF$_O$ in Table 1), for a 
good reason: the Ohmic diffusivity enables the magnetic flux 
that would have gone into the central object in the ideal MHD 
limit to accumulate in a small circumstellar region where the 
magnetic braking is particularly efficient (see Fig.\ \ref{PureO}), 
as in the AD case. Such a magnetic flux accumulation was not 
obvious in \citeauthor{Machida_2010}'s simulations, which may again indicate 
an enhanced magnetic diffusion, either of numerical origin or 
through 3D effects that are not captured by our 2D calculations.

\subsubsection{Weak Core Magnetization and RSD Formation} 

Here we comment on the possibility that dense star forming cores 
may be magnetized to different levels, and disks form preferentially 
in those that are weakly magnetized. Our calculations indicate that, 
in the presence of ambipolar diffusion, the core mass-to-flux ratio 
need to be greater than at least $\sim 10$ in order for a rotationally 
supported disk to form and survive to late times. Observationally, a 
mean value of $\lambda_{los} \approx 4.8 \pm 0.4$ is inferred by 
\citet{TrolandCrutcher2008} from the line-of-sight field strength for 
a sample of dark cloud cores. Applying geometric corrections would 
reduce the value statistically by a factor of 2--3 \citep{Shu_1999}, 
making it unlikely for the majority of dense cores to be magnetized 
as weakly as $\lambda \gtrsim 10$. Since the majority of, if not all, 
young stars (formed out of all dense cores) are thought to have an 
RSD at some point, we consider it unlikely that weak core magnetization 
is the main reason for RSD formation. 

\subsection{Observational Implications: Disk vs Pseudodisk}

We should emphasize that, in our simulations, even though large 
rotationally supported disks are difficult to form, highly 
flattened dense ``disk-like'' structures are prevalent (see  
Figs.\ \ref{ReferenceModel}, \ref{PureO}, \ref{WeakB_NoDisk}, 
and \ref{PureHall_vphi}). This is not surprising because, just 
like rotation, the magnetic field can provide anisotropic support 
to the cloud core, allowing matter to settle along field lines 
into flattened structures \citep{GalliShu1993}. The fact that 
there are two types of forces in nature that can retard
(anisotropically) the gravitational collapse naturally leads 
to two types of flattened structures: rotationally supported
disks and magnetically induced (pseudo-)disks. If the dense
cores are as strongly magnetized as indicated by the currently
available observations (with a dimensionless mass-to-flux 
ratio of several or smaller, see \S{}\ref{intro}), then there is 
typically more magnetic energy than rotational energy, and a 
magnetically induced pseudo-disk is just as, if not more, likely 
to form around an accreting protostar as a rotationally 
supported disk, especially in view of the fact that magnetic braking 
hinders the formation of rotationally supported disks but not 
pseudodisks. It is therefore premature to conclude that dense 
flattened structures observed around deeply embedded protostars 
(such as from dust continuum observations, e.g., \ct{Jorgensen_2009})
are rotationally supported disks rather than magnetically 
induced pseudodisks; the latter can be just as thin as (perhaps 
even thinner than) rotationally supported disks, because of magnetic 
compression. To confuse the situation further, the pseudodisks can 
have a substantial rotation as well (just not enough to provide 
the full support against gravity) and may or may not collapse 
at a high speed (see again Figs.\ \ref{ReferenceModel}, 
\ref{PureO}, \ref{WeakB_NoDisk} and \ref{PureHall_vphi}). 
Detailed kinematic information, as well as a knowledge of 
the central mass, are needed to establish whether a dense 
flattened circumstellar structure is a rotationally supported 
disk or not. High resolution observations of the circumstellar 
magnetic field structure, such as those in \citet{Girart_2006}, 
will also go a long way towards testing the idea of magnetically  
induced pseudodisk \citep{Goncalves_2008}.

\section{Summary}
\label{summary}

We have carried out a set of 2D axisymmetric calculations exploring
non-ideal MHD effects in magnetic braking and protostellar disk 
formation in rotating magnetized dense cores. Our main conclusions 
are summarized as follows:

1. For a realistic magnetic field of moderate strength corresponding to a core
   mass-to-flux ratio $\lambda \sim$ 3--4, the magnetic braking is
   strong enough to remove essentially all of the angular momentum 
   of the material that accretes onto the central object in the
   presence of ambipolar diffusion under a wide range of conditions in 
   2D. Any large-scale (greater than
   several AUs) rotationally supported disk (RSD) is suppressed by the
   formation of an ambipolar diffusion-induced accretion shock, which
   traps a strong magnetic field near the central object, leading to
   efficient magnetic braking of the post-shock material. 

2. On scales greater than $\sim 10\AU$, realistic levels of Ohmic 
   diffusivity do not 
   enable the formation of large-scale RSDs, either by itself or in 
   combination with ambipolar diffusion. Furthermore, Ohmic 
   dissipation does not 
   necessarily 
   reduce the magnetic braking efficiency. It can make the braking 
   more efficient by enabling magnetic flux accumulation 
   at small radii, where the field strength is increased, similar 
to the case of ambipolar diffusion. 

3. The Hall effect can spin up the post-AD shock material to a 
   significant, supersonic rotation speed, although the rotation 
   remains too sub-Keplerian to form an RSD for the parameter space 
explored in this work.
 
4. For an unusually weak magnetic field corresponding to a core
   mass-to-flux ratio $\lambda \gtrsim 10$, a small RSD often forms early
   in the protostellar accretion phase, when the central mass is still
   small. In the majority of cases, the RSD disappears at later times,
   braked strongly by the powerful outflow that it drives. In some
   cases, particularly when the cosmic ray ionization rate is unusually 
low and the core rotation rate is unusually high, the fate of the 
early disk is unknown
   because the simulation stops early due to numerical difficulty. 

5. We discussed several possible ways to enable the formation of
large-scale RSDs: magnetic instabilities in 3D, early formation 
of small RSDs at high densities, outflow stripping of protostellar 
envelope, enhanced magnetic diffusion and weak core magnetization.  
The more likely of these possibilities are, in our view, the 
weakening AD shock in 3D through interchange instability, which 
is expected to decrease the field strength (and thus the braking 
efficiency) near the central object, outflow stripping of 
protostellar envelope, which may allow rapid formation of a 
large-scale RSD during the transition from the deeply embedded 
(Class 0) phase to more revealed (Class I and II) phase of low-mass 
star formation, and enhanced magnetic diffusivity, which may be 
driven by turbulence-induced reconnections. 

\acknowledgments
This work was supported in part by NASA through NNG06GJ33G and
NNX10AH30G, by the Theoretical Institute for Advanced Research 
in Astrophysics (TIARA) through the CHARMS initiative, and 
by the National Science Council of Taiwan through grant 
NSC97-2112-M-001-018-MY3.


\begin{thebibliography}{}
\bibitem[Allen et al.(2003)Allen, Li, \& Shu]{Allen_2003}Allen, A., Li, Z.-Y., \& Shu, F. H. 2003, \apj, 599, 363 
\bibitem[Alves et al.(2007)]{Alves_2007}Alves, J., Lombardi, M., \& Lada, C. J. 2007, \aap, 462, L17
\bibitem[Andr{\'e} et al.(2010)]{Andre_2010}Andr{\'e}, P., et al.\ 2010, \aap, 518, L102 
\bibitem[Arce \& Sargent(2006)]{ArceSargent2006}Arce, H. G., \& Sargent, A. I. 2006, \apj, 646, 1070 
\bibitem[Basu \& Mouschovias(1994)]{BasuMouschovias1994} Basu, S., \& Mouschovias, T. C. 1994, \apj, 432, 720 
\bibitem[Bergin \& Tafalla(2007)]{BerginTafalla2007}Bergin, E. A., \& Tafalla, M. 2007, \araa, 45, 339
\bibitem[Bodenheimer(1995)]{Bodenheimer1995}Bodenheimer, P. 1995, \araa, 33, 199
\bibitem[Bontemps et al.(1996)]{Bontemps_1996}Bontemps, S., Andr{\'e}, P., Terebey, S., \& Cabrit, S. 1996, \aap, 311, 858 
\bibitem[Boss(1998)]{Boss1998} Boss, A. P. 1998, Origins, 148, 314 
\bibitem[C. Braiding(2011)]{Braiding2011} Braiding, C. 2011, unpublished PhD thesis, Macquarie University
\bibitem[Ciolek \& K{\"o}nigl(1998)]{CiolekKonigl1998} Ciolek, G. E., \& K{\"o}nigl, A. 1998, \apj, 504, 257 
\bibitem[Clarke et al.(1994)]{Clarke_1994} Clarke, D. A., Norman, M. L., \& Fiedler, R. A. 1994,
ZEUS-3D User Manual (Tech.\ Rep.\ 015; Urbana-Champaign: National Center for Supercomputing Applications)
\bibitem[Contopoulos et al.(1998)]{Contopoulos_1998} Contopoulos, I., Ciolek, G. E., \& K{\"o}nigl, A. 1998, \apj, 504, 247
\bibitem[Crutcher et al.(2010)]{Crutcher_2010}Crutcher, R. M., Wandelt, B., Heiles, C., Falgarone, E., \& Troland, T. H. 2010, \apj, 725, 466
\bibitem[Dapp \& Basu(2010)]{DappBasu2010}Dapp, W. B., \& Basu, S. 2010, \aap, 521, 56
\bibitem[Dib et al.(2007)]{Dib_2007}Dib, S., Kim, J., V{\'a}zquez-Semadeni, E., Burkert, A., \& Shadmehri, M. 2007, \apj, 661, 262 
\bibitem[Duffin \& Pudritz(2009)]{DuffinPudritz2009}Duffin, D. F. \& Pudritz, R. E. 2009, \apj, 706, L46
\bibitem[Fleming et al.(2000)]{Fleming_2000}Fleming, T., P., Stone, J. M. \& Hawley, J. F. 2000, \apj, 530, 464 
\bibitem[Galli et al.(2006)]{Galli_2006} Galli, D., Lizano, S., Shu, F. H., \& Allen, A. 2006, \apj, 647, 374 
\bibitem[Galli \& Shu(1993)]{GalliShu1993} Galli, D., \& Shu, F. H. 1993, \apj, 417, 243 
\bibitem[Girart et al.(2006)Girart, Rao, \& Marrone]{Girart_2006} Girart, J. M., Rao, R., \& Marrone, D. P. 2006, Science, 313, 812 
\bibitem[Gon{\c c}alves et al.(2008)]{Goncalves_2008}Gon{\c c}alves, J., Galli, D., \& Girart, J. M. 2008, \aap, 490, L39
\bibitem[Goodman et al.(1993)]{Goodman_1993}Goodman, A. A., Benson, P. J., Fuller, G. A., \& Myers, P. C. 1993, \apj, 406, 528 
\bibitem[Heiles \& Troland(2005)]{HeilesTroland2005}Heiles, C., \& Troland, T. H. 2005, \apj, 624, 773 
\bibitem[Hennebelle \& Ciardi(2009)]{HennebelleCiardi2009}Hennebelle, P., \& Ciardi, A. 2009, \aap, 506, L29
\bibitem[Hennebelle et al.(2011)]{Hennebelle_2011}Hennebelle, P., Commer{\c c}on, B., Joos, M., Klessen, R. S., Krumholz, M., Tan, J. C., \& Teyssier, R. 2011, \aap, 528, A72
\bibitem[Hennebelle \& Fromang(2008)]{HennebelleFromang2008}Hennebelle, P., \& Fromang, S. 2008, \aap, 477, 9
\bibitem[Hosking \& Whitworth(2004)]{HoskingWhitworth2004}Hosking, J. G. \& Whitworth, A. P. 2004, \mnras, 347, 1001
\bibitem[Huba(2003)]{Huba2003} Huba, J. D. 2003, in Space Plasma Simulation, ed.\ J. B{\"u}chner, C. Dum, \& M. Scholer, Lecture Notes in Physics, vol.\ 615 (Berlin: Springer), 166
\bibitem[J{\o}rgensen et al.(2009)]{Jorgensen_2009}J{\o}rgensen, J. K., van Dishoeck, E. F., Visser, R., Bourke, T. L.,
Wilner, D. J., Lommen, D., Hogerheijde, M. R., \& Myers, P. C. 2009, \aap, 507, 861
\bibitem[K{\"o}nigl \& Pudritz(2000)]{KoniglPudritz2000}K{\"o}nigl, A. \& Pudritz, R. 2000, in Protostars and Planets IV, eds.\ V. Mannings et al.\ (Univ.\ of Arizona Press), 759 
\bibitem[K{\"o}nigl et al.(2010)]{Konigl_2010}K{\"o}nigl, A., Salmeron, R. \& Wardle, M. 2010, \mnras, 401, 479
\bibitem[Kowal et al.(2009)]{Kowal_2009}Kowal, G., Lazarian, A., Vishniac, E. T., \& Otmianowska-Mazur, K. 2009, \apj, 700, 63 
\bibitem[Kudoh \& Basu(2011)]{KudohBasu2011} Kudoh, T., \& Basu, S. 2011, \apj, 728, 123 
\bibitem[Kunz \& Mouschovias(2010)]{KunzMouschovias2010}Kunz, M. W., \& Mouschovias, T. C. 2010, \mnras, 408, 322 
\bibitem[Krasnopolsky \& K{\"o}nigl(2002)]{KrasnopolskyKonigl2002}Krasnopolsky, R., K{\"o}nigl, A. 2002, \apj, 580, 987 
\bibitem[Krasnopolsky et al.(2010)]{Krasnopolsky_2010}Krasnopolsky, R., Li, Z.-Y. \& Shang, H. 2010, \apj, 716, 1541
\bibitem[Krasnopolsky et al.(2011)]{Krasnopolsky_2011}Krasnopolsky, R., Li, Z.-Y. \& Shang, H. 2011, \apj, in press
\bibitem[Li \& McKee(1996)]{LiMcKee1996} Li, Z.-Y., \& McKee, C. F. 1996, \apj, 464, 373 
\bibitem[Lizano \& Shu(1989)]{LizanoShu1989} Lizano, S., \& Shu, F. H. 1989, \apj, 342, 834 
\bibitem[Mac Low et al.(1995)]{MacLow_1995} Mac Low, M.-M., Norman, M. L., K{\"o}nigl, A., \& Wardle, M. 1995, \apj, 442, 726 
\bibitem[Machida et al.(2007)]{Machida_2007}Machida, M. N., Inutsuka, S., \& Matsumoto, T. 2007, \apj, 670, 1198 
\bibitem[Machida et al.(2010)]{Machida_2010}Machida, M. N., Inutsuka, S., \& Matsumoto, T. 2010, arXiv:1009.2140
\bibitem[Masunaga \& Inutsuka(2000)]{MasunagaInutsuka2000}Masunaga, H., \& Inutsuka, S. 2000, \apj, 531, 350
\bibitem[Matzner \& McKee(2000)]{MatznerMcKee2000}Matzner, C. D. \& McKee, C. F. 2000, \apj, 545, 364
\bibitem[Mathis et al.(1977)]{Mathis_1977}Mathis, J. S., Rumpl, W., \& Nordsieck, K. H. 1977, \apj, 217, 425
\bibitem[Mellon \& Li(2008)]{MellonLi2008}Mellon, R. R., \& Li, Z.-Y. 2008, \apj, 681, 1356
\bibitem[Mellon \& Li(2009)]{MellonLi2009}Mellon, R. R., \& Li, Z.-Y. 2009, \apj, 698, 922
\bibitem[Nakamura \& Li(2005)]{NakamuraLi2005}Nakamura, F., \& Li, Z.-Y. 2005, \apj, 631, 411
\bibitem[Nakano \& Nakamura(1978)]{NakanoNakamura1978} Nakano, T., \& Nakamura, T. 1978, \pasj, 30, 671 
\bibitem[Nakano et al.(2002)Nakano, Nishi, \& Umebayashi]{Nakano_2002} Nakano, T., Nishi, R., \& Umebayashi, T. 2002, \apj, 573, 199
\bibitem[Nishi et al.(1991)]{Nishi_1991}Nishi, R., Nakano, T., \& Umebayashi, T. 1991, \apj, 368, 181
\bibitem[Norman \& Heyvaerts(1985)]{NormanHeyvaerts1985}Norman, C. A., \& Heyvaerts, J. 1985, \apj, 147, 247
\bibitem[Pagani et al.(2010)]{Pagani_2010}Pagani, L., Steinacker, J., Bacmann, A., Stutz, A., \& Henning, T. 2010, Science, 329, 1622 
\bibitem[Padovani et al.(2009)]{Padovani_2009}Padovani, M., Galli, D. \& Glassgold, A. E. 2009, \aap, 501, 619
\bibitem[Price \& Bate(2007)]{PriceBate2007} Price, D. J., \& Bate, M. R. 2007, \apss, 311, 75 
\bibitem[Sano \& Stone(2002)]{SanoStone2002}Sano, T., \& Stone, J. M. 2002, \apj, 570, 314
\bibitem[Santos-Lima et al.(2010)]{Santos-Lima_2010}Santos-Lima, R., Lazarian, A., de Gouveia Dal Pino, E. M., \& Cho, J. 2010, \apj, 714, 442 
\bibitem[Shang et al.(2006)]{Shang_2006}Shang, H., Allen, A., Li, Z.-Y., Liu, C.-F., Chou, M.-Y., \& Anderson, J.\ 2006, \apj, 649, 845 
\bibitem[Shu et al.(1999)]{Shu_1999}Shu, F. H., Allen, A., Shang, H., Ostriker, E. C. \& Li, Z.-Y. 1999, in The Origins of Stars and Planetary Systems,  eds.\ C. J. Lada \& N. D. Kylafis (Kluwer), 193
\bibitem[Shu et al.(2006)]{Shu_2006} Shu, F. H., Galli, D., Lizano, S., \& Cai, M. 2006, \apj, 647, 382 
\bibitem[Shu \& Li(1997)]{ShuLi1997}Shu, F. H. \& Li, Z.-Y. 1997, \apj, 475, 251
\bibitem[Shu et al.(2000)]{Shu_2000}Shu, F. H., Najita, J. Shang, H. \& Li, Z.-Y. 2000, in Protostars and Planets IV, eds.\ V. Mannings et al.\ (Univ.\ of Arizona Press), 789
\bibitem[Shu et al.(1991)]{Shu_1991}Shu, F. H., Ruden, S. P., Lada, C. J. \& Lizano, S. 1991, \apj, 370, 31
\bibitem[Tafalla et al.(1998)]{Tafalla_1998}Tafalla, M., Mardones, D., Myers, P. C., Caselli, P., Bachiller, R., \& Benson, P. J. 1998, \apj, 504, 900 
\bibitem[Tassis \& Mouschovias(2007)]{TassisMouschovias2007} Tassis, K., \& Mouschovias, T. C. 2007, \apj, 660, 388 
\bibitem[Tilley \& Pudritz(2005)]{TilleyPudritz2005} Tilley, D. A., \& Pudritz, R. E. 2005, Protostars and Planets V, 8473 
\bibitem[Troland \& Crutcher(2008)]{TrolandCrutcher2008} Troland, T. H., \& Crutcher, R. M. 2008, \apj, 680, 457 
\bibitem[Ward-Thompson et al.(2000)]{Ward-Thompson_2000}Ward-Thompson, D., Kirk, J. M., Crutcher, R. M., Greaves, J. S., Holland, W. S., \& Andr{\'e}, P. 2000, \apjl, 537, L135 
\bibitem[Ward-Thompson et al.(1999)]{Ward-Thompson_1999}Ward-Thompson, D., Motte, F., \& Andr{\'e}, P.\ 1999, \mnras, 305, 143 
\bibitem[Wardle \& K{\"o}nigl(1993)]{WardleKonigl1993}Wardle, M., \& K{\"o}nigl, A. 1993, \apj, 410, 218
\bibitem[Wardle \& Ng(1999)]{WardleNg1999} Wardle, M., \& Ng, C. 1999, \mnras, 303, 239
\bibitem[Webber(1998)]{Webber1998}Webber, W. R. 1998, \apj, 506, 329
\bibitem[Williams \& Cieza(2011)]{WilliamsCieza2011}Williams, J. P., \& Cieza, L. A. 2011, \araa, in press
\end{thebibliography}
\end{document}